\documentclass[twocolumn]{aastex62}
\usepackage{graphics}
\usepackage{multirow}
\usepackage{rotating}
\usepackage{amsfonts}
\usepackage{amsmath}
\usepackage{enumerate}
\usepackage{tabularx}
\usepackage{natbib,amsmath,amssymb,epsfig,graphicx,longtable,multirow,float,url,footnote}
\usepackage{lineno}
\usepackage{amssymb}
\usepackage{morefloats}
\usepackage{listings}

\usepackage{url}

\usepackage{epstopdf}

\usepackage[parfill]{parskip}    

\newcommand{\prs}{$^{\prime\prime}\;$}
\newcommand{\pr}{$^{\prime\prime}$}

\received{}
\revised{}
\accepted{}
\submitjournal{AJ}

\begin{document}

\setcounter{table}{0}
\setcounter{figure}{0}


\title{First ALMA Millimeter Wavelength Maps of Jupiter, with a Multi-Wavelength Study of Convection}




\correspondingauthor{Imke de Pater}
\email{imke@berkeley.edu}

\author[0000-0002-4278-3168]{Imke de Pater}
\affiliation{ Department of Astronomy \\
  501 Campbell Hall \\
  University of California \\
  Berkeley, CA 94720, USA}

\author[0000-0001-9209-7716]{R. J. Sault}
\affiliation{ School of Physics\\
  University of Melbourne\\
  Victoria, 3010, Australia.}

\author[0000-0002-6293-1797]{Chris Moeckel}
\affiliation{Department of Earth and Planetary Sciences\\
  McCone Hall \\
  University of California\\
  Berkeley, CA 94720, USA}

\author[0000-0002-9820-1032]{Arielle Moullet}
\affiliation{SOFIA/USRA\\
  NASA Ames Building N232\\
  Moffett Field, CA 94035, USA}

\author[0000-0003-2804-5086]{Michael H. Wong}
\affiliation{ Department of Astronomy \\
  501 Campbell Hall \\
  University of California \\
  Berkeley, CA 94720, USA}

\author[0000-0001-8890-428X]{Charles Goullaud}
\affiliation{ Department of Astronomy \\
  501 Campbell Hall \\
  University of California \\
  Berkeley, CA 94720, USA}

\author[0000-0003-3197-2294]{David DeBoer}
\affiliation{ Department of Astronomy \\
  501 Campbell Hall \\
  University of California \\
  Berkeley, CA 94720, USA}

\author[0000-0002-5344-820X]{Bryan Butler}
\affiliation{National Radio Astronomy Observatory\\
  Socorro, NM 87801, USA.}

\author[0000-0002-9679-4153]{Gordon Bjoraker}
\affiliation{Goddard Space Flight Center\\
  8800 Greenbelt Rd, Greenbelt MD 20771, USA}

\author[0000-0003-1869-0938]{M\'at\'e \'Ad\'amkovics}
\affiliation{Dept. of Physics \& Astronomy\\
  Clemson University\\
  Clemson, SC 29634-0978, USA}

\author[0000-0003-3047-615X]{Richard Cosentino}
\affiliation{Goddard Space Flight Center\\
  8800 Greenbelt Rd, Greenbelt MD 20771, USA}

\author[0000-0002-4241-0302]{Padraig T. Donnelly}
\affiliation{ Department of Physics and Astronomy\\
  University of Leicester\\
  University Road, Leicester, LE1 7RH, UK}
  
\author[0000-0001-5834-9588]{Leigh N. Fletcher}
\affiliation{ Department of Physics and Astronomy\\
  University of Leicester\\
  University Road, Leicester, LE1 7RH, UK}

\author[0000-0002-8160-3553]{Yasumasa Kasaba}
\affiliation{ Planetary Plasma and Atmospheric Research Center (PPARC)\\
  Tohoku University, Japan}

\author[0000-0001-7871-2823]{Glenn Orton}
\affiliation{Jet Propulsion Laboratory\\
  California Institute of Technology\\
  4800 Oak Grove Dr, Pasadena, CA 91109, USA}

\author[0000-0002-4239-5907]{John Rogers}
\affiliation{British Astronomical Association\\ 
  Burlington House, Piccadilly, London  W1J 0DU, KU}

\author[0000-0001-5374-4028]{James Sinclair}
\affiliation{Jet Propulsion Laboratory\\
  California Institute of Technology\\
  4800 Oak Grove Dr, Pasadena, CA 91109, USA}

\author[0000-0003-4314-4947]{Eric Villard}
\affiliation{Joint ALMA Observatory/ESO\\
  Avenida Alonso de Cordova 3107, Vitacura, Santiago, Chile}

\vspace{2.0in}

\pagebreak






\pagebreak

\begin{abstract}

We obtained the first maps of Jupiter at 1--3 mm wavelength with the
Atacama Large Millimeter/Submillimeter Array (ALMA) on 3--5 January
2017, just days after an energetic eruption at 16.5$^\circ$S
jovigraphic latitude had been reported by the amateur community, and
about 2-3 months after the detection of similarly energetic eruptions
in the northern hemisphere, at 22.2--23.0$^\circ$N. Our observations,
probing below the ammonia cloud deck, show that the erupting plumes in
the SEB bring up ammonia gas from the deep atmosphere. 
While models of plume eruptions that are triggered at the water
condensation level explain data taken at uv--visible and mid-infrared
wavelengths, our ALMA observations provide a crucial, hitherto missing, link
in the moist convection theory by showing that ammonia gas from the
deep atmosphere is indeed brought up in these plumes. Contemporaneous
HST data show that the plumes reach altitudes as high as the tropopause.
We suggest that the plumes at 22.2--23.0$^\circ$N also rise up well above the ammonia cloud deck, and that descending air may dry the neighboring belts even more than in quiescent times, which would explain our observations in the north. 

\end{abstract}

\bigskip

\keywords{Jupiter --- Atmosphere --- Radio observations --- Radiative Transfer}

\pagebreak

\section{Introduction}\label{sec:intro}

Numerous ground-based and space-borne telescopes have monitored
Jupiter closely during the past few years, being motivated to provide
support to NASA's {\it Juno} mission, in particular during close encounters
of the spacecraft with Jupiter, referred to as Perijoves
(PJs). Although Juno data are not included in this paper, the
observations discussed were similarly motivated. They were carried out 
in early January 2017,  near {\it Juno}'s originally planned PJ8
(which was 11 Jan. 2017). Contributing uniquely to this campaign, observations were obtained
with the Atacama Large Millimeter/Submillimeter Array (ALMA). This is the first time
that ALMA observed Jupiter's atmosphere at 1.3 and 3 mm (233 and 97
GHz), probing 40-50 km below the visible ammonia-ice cloud (down to
3-4 bar). Data at these wavelengths complement the Very Large Array
(VLA) Jupiter maps of 2013-2014 in the cm wavelength range (de
Pater et al., 2016, 2019; henceforth dP16 and dP19, respectively).

\begin{figure*}
\includegraphics[width=32pc]{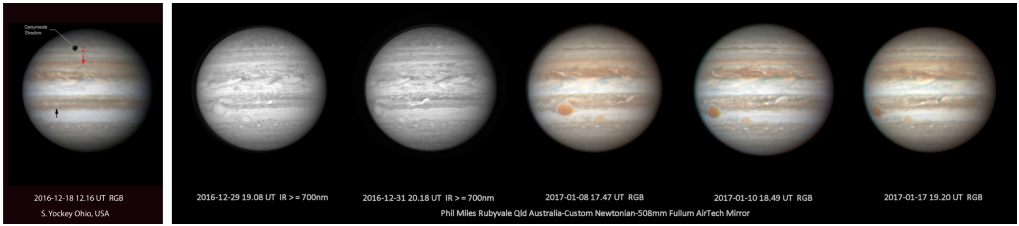}
\caption{Images in visible light spanning the origin of the SEB outbreak.  A pre-outbreak image
on 28 Dec. 2016; discovery image on 29 Dec. 2016; and subsequent images on 31 Dec. 2016, and 8, 10 and 17 Jan. 2017.
The source of the Outbreak remains fixed near 208$^\circ$ System II, and the disturbance propagates to the east, as shown by the sequence of images. (Courtesy S. Yockey, Ohio, USA; Phil Miles, Australia). 
}
\label{fig:S1}\end{figure*}

Fortuitously, the timing of the ALMA observations was just a few days
after amateur astronomer Phil Miles announced the onset of an
``outbreak'' in Jupiter's South Equatorial Belt (SEB;
7--20$^\circ$S\footnote{All latitudes are referred to as
  planetographic latitudes.}): a small bright white plume at
16.5$^\circ$S that signified the start of a large-scale disruption in
the SEB (Fig.~\ref{fig:S1}). The last full fade and revival cycle of the SEB took
place in 2009-2011 (Fletcher et al., 2011; 2017a), where the word
``fading'' is used when the SEB looses its brown color and turns white
(like a lighter-colored axisymmetric band, referred to as a
``zone''). Although the present outbreak was not preceded by a period of
fading, there are many similarities between this outbreak and the
revival cycle following the 2009-2011 fade, as shown in this paper. While outbreaks in the SEB occur at irregular intervals of a few years, periods between faded states can be over three decades long (Rogers, 1995; Fletcher, 2017).

Meanwhile, in the northern hemisphere, three months prior to our
observations, four extremely bright white plumes had been discovered
at 22.2--23.0$^\circ$N, i.e., just south of the North Temperate
Belt (NTB; 24--31$^\circ$N). Over the next few months this led to a
planetary-scale disturbance in the NTB, resulting in a uniform orange belt by the end of November 2016, at latitudes spanning 22.8--26.7$^\circ$Nke (S\'anchez-Lavega et. al., 2017).  
Such NTB outbreaks occur on timescales $\gtrsim$5 years.

Radio observations at mm--cm wavelengths are unique because they probe below the visible cloud deck (dP19). 
Therefore, our ALMA data give a unique perspective on the SEB outbreak and the aftermath of the NTB revival since these are the only data that let us trace these events below the ammonia cloud deck. In the case of the NTB, our data were acquired after the entire belt had ``revived'', but in case of the SEB the data were taken during the period when plume eruptions were in progress. 

We present the observations in Section 2, the results in Section 3
with models in Section 4, concluding with a discussion and a possible
explanation in the context of moist convection theory in Section 5. A
brief summary is provided in Section 6.

\section{OBSERVATIONS}

Jupiter was observed with ALMA on 3-5 January 2017, when the array was
composed of 40 antennas, and placed in a relatively compact
configuration (C40-2). Observations were obtained in Band 3 (3 mm,
90--105 GHz) and Band 6 (1.3 mm, 223--243 GHz).
The observations are summarized in Table 1.

Quasi-simultaneous observations were obtained at several other
telescopes on January 10--14. Specifically, observations at a spatial
resolution 3.5--4 times higher than that of the 1.3 mm ALMA data were
obtained with the Very Large Array (VLA) in the X-band ($\sim$3.5 cm,
8--12 GHz); although the spatial resolution in these maps is
exquisite, the large scale structure is poorly mapped. We used the
Hubble Space Telescope (HST) WFC2/UVIS camera at multiple wavelengths
to map the visible cloud structure, including bright plumes. With the
Gemini telescope we imaged the planet at a wavelength of 5 $\mu$m
using the NIRI instrument, while we simultaneously obtained 5-$\mu$m
spectroscopic data with the Keck telescope using the NIRSPEC
spectrometer, both probing down to 7--8 bar in cloud-free regions. To
diagnose thermal effects of the SEB outbreak on the upper troposphere
and stratosphere, we used mid-infrared detectors on the Very Large
Telescope (VLT), VISIR, and Subaru telescope, COMICS. Table 2 provides
a summary of all observations taken in addition to the ALMA data. In
the following subsections we describe each of the observations in more
detail.

\begin{figure*}
\includegraphics[scale=0.9]{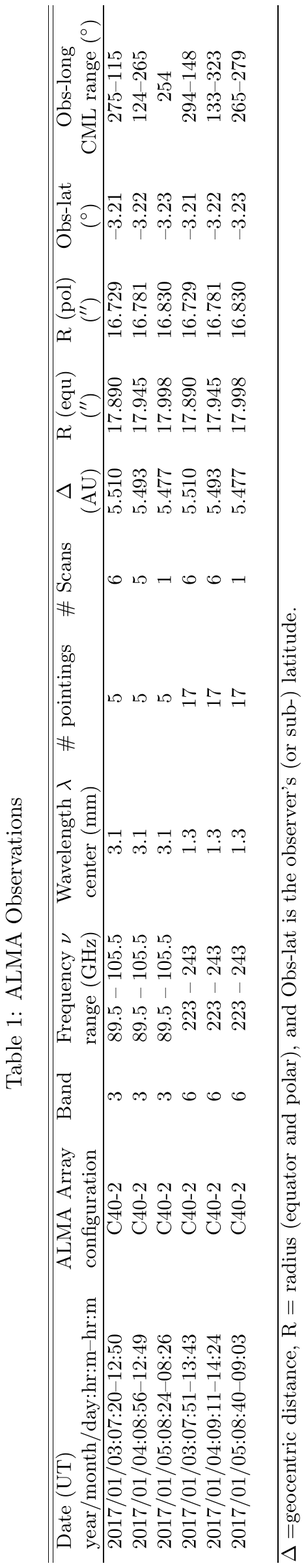}
\end{figure*}

\begin{figure*}
\includegraphics[scale=0.9]{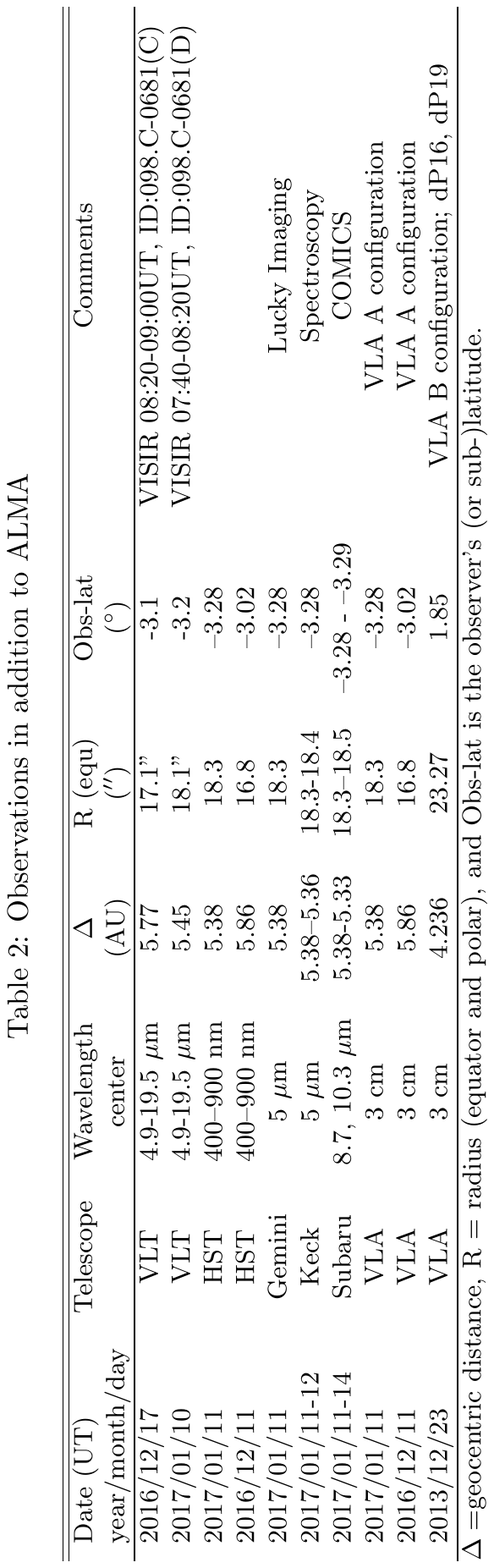}
\end{figure*}

\subsection{ALMA}

We obtained 5 -- 6 observations (or ``executions'') with ALMA (program
2016.1.00701.S) on each of the first two days (3 and 4 Jan. 2017),
interleaving Band 3 (3 mm, 90--105 GHz) and Band 6 (1.3 mm, 223--243
GHz); one additional observation was taken on 5 Jan. Since Jupiter is large,
roughly 35\prs across during the observing period, we used the
mosaicing method to map the entire planet. A total of 5 pointings were
used in Band 3, and 17 in Band 6, so that the majority of time was
spent in Band 6. In each setup, we had 4 spectral windows, each 2 GHz
wide.

The basic data received from an interferometer array, such as the VLA
or ALMA,
are (complex) visibilities, formed by correlating signals from the
array's elements. These are measured in the u-v plane, where the
coordinates u and v describe the separation, or baseline, between two
antennas (i.e., an interferometer) in wavelength, as projected on the
sky in the direction of the source. We refer the reader to 
de Pater et. al. (2019) for a summary of this technique. 

The initial flagging and calibration was done using the ALMA pipeline in
the Common Astronomy Software Applications package, CASA.
Unfortunately, the absolute flux density of Jupiter in the
various observations was obtained using different flux calibrators,
which resulted in slightly different flux scales between
executions. For all observations J1256-0547 was used as phase
calibrator. We modified the flux densities so that all scans were
referenced to Callisto, for which we used the internal model in CASA
(Butler-Horizons
2012\footnote{(https://science.nrao.edu/facilities/alma/aboutALMA/Technology/ALMA$_-$Memo$_-$Series/alma594/memo594.pdf}). We
modified the phases of Jupiter to take out its motion across the
sky. The MIRIAD software package (Sault et al., 1995) was used to
create maps of the planet\footnote{The CASA software package at the
  time did not produce reliable mosaicked images; this has been remedied in CASA 5.4.0 (NAASC$_-$117).} 
.
ALMA's primary beam was assumed to be a gaussian with FWHM $1.13\lambda/D$ radians ($\lambda =$ wavelength; $D=$ diameter ALMA dish, which we assumed to be 12 m). 
The procedures to produce longitude-smeared and longitude-resolved
maps were then similar to those used in earlier VLA observations, including
self-calibration (Sault et al., 2004; dP16, dP19). However, the
techniques were generalized to account for beam effects and mosaicking.
Due to the excellent u-v-coverage in ALMA data compared to the VLA, the maps are essentially devoid of instrumental artefacts.

As in the previous papers, in order to best assess small variations on Jupiter's disk, a
limb-darkened disk was subtracted from the u-v data with a brightness
temperature and limb-darkening parameter that produced a best fit (by eye) to
the data (i.e., ``best fit'' means parameters such that there is no planet after imaging the residual u-v data). Limb-darkening was modeled by multiplying the brightness
temperature at disk center, T$_b'$, by (cos$\theta$)$^q$, with $\theta$ the emission
angle on the disk (i.e., the angle between the surface normal vector
and the line-of-sight vector to Earth), and $q$ a constant that
provides a best fit to the data. Although more complex limb-darkening
models could be used instead of our simple algorithm, our main goal is
to subtract the large bright smoothly-varying structure that is
Jupiter's disk, so we can produce reliable maps of the residuals. The
subtracted disk is added back before we model the data with radiative
transfer calculations (see also dP19).

Disks which provided a best-fit to the Band 3  (T$_b' =$ 131 K
with $q=$ 0.10) and Band 6 (T$_b' =$ 115 K with $q=$ 0.08) data revealed
brightness temperatures that were only of order 60--70\% of what we expected.
Although our data lacked short
spacings, (in Cycle 4 it was not possible to simultaneously use
the Atacama Compact Array (ACA) and the 12-m array),  
this was not the reason for the low observed brightness
temperatures. These appear to be caused by errors in the ALMA
observations and pipeline
reduction software. Based on an ALMA memo on
calibration\footnote{http://library.nrao.edu/public/memos/alma/main/memo318.pdf},
we conclude that the system temperature, Tsys, is usually determined
on blank sky. This is reasonable for a source which does not contribute significantly to Tsys. However this approach is not appropriate for very bright sources. For example for ALMA observations of the Sun, Tsys is determined on the disk of the Sun\footnote{https://almascience.nrao.edu/alma-data/science-verification/sunspot-calibration}. A similar approach should be used when observing the bright planets as well.

 In order to remedy this shortcoming, 
we assumed disk-averaged brightness
temperatures based on the best model fits to dP19's disk-averaged
brightness temperature spectrum (Fig. 4 in dP19, with the Karim et
al. (2018) model), and scaled the data accordingly. The values used are listed in Table 3,
T$_b$(adopt). We then calculated the brightness temperature at disk
center, T$_b$(cent), that would provide T$_b$(adopt) when using the
limb-darkening parameter $q$, and after subtracting the cosmic
microwave background (T$_{\rm cmb}$) to mimic the observations. After
putting the originally subtracted disk back, we multiplied the maps by
T$_b$(cent)/T$_b'$ and added T$_{\rm cmb}$ to match the observations
as close as possible to Jupiter's disk.

\begin{figure*}
\includegraphics[scale=0.9]{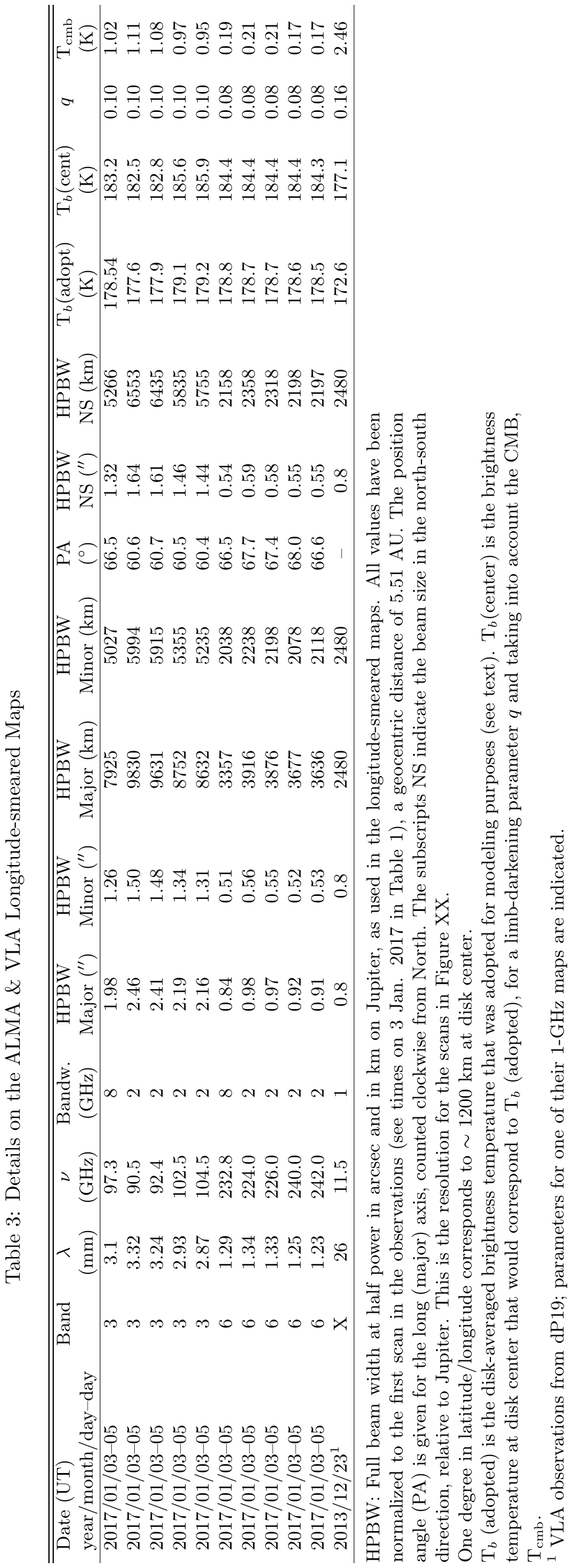}
\end{figure*}

\subsection{VLA}

Jupiter was observed with the VLA (program 16B-048) on 11 Dec. 2016 and 11 Jan. 2017. Observations were obtained in the X band (8--10 GHz in 2016; 8--12 GHz in 2017), while the VLA was in its most extended (A) configuration. 
The VLA data were processed using the standard pipeline; the data were
averaged in time (10s) and in frequency (8 channels) and remaining
noisy baselines were removed manually. The dataset was reduced using
the full 2 or 4 GHz bandwidth and additionally, the data were split
spectrally into 1-GHz wide sets and each set was self-calibrated once
on a limb-darkened model with brightness temperature and
limb-darkening coefficient obtained from dP16.  After subtracting the
aforementioned model, longitude-resolved images were formed using the MIRIAD software package (Sault et al., 2004; dP16, dP19).  
Details on the observations are provided in Tables 2 and 3.

\subsection{HST}

Images in the UV/visible/near-IR range were taken on 11 Dec. 2016 (program GO-14661) and 11 Jan. 2017 (program GO-14839 ) with the UVIS detector of the WFC3 instrument aboard the Hubble Space Telescope (Dressel 2019; see their Table 6.2 for filter properties). Raw data are available from the Hubble MAST archive, and processed data are available from https://archive.stsci.edu/prepds/wfcj.

Corrections were applied for fringing at long wavelengths (Wong, 2011), and cosmic ray hits were removed based on their sharpness (van Dokkum, 2001). The images were navigated by aligning the data to a synthetic limb-darkened disk, as described in Lii et al. (2010).

\subsection{VLT}

Thermal-infrared observations in eight narrow-band filters between
7-20 $\mu$m were acquired by the VISIR instrument, with a 9$^{th}$
band covering the 5-$\mu$m window (Lagage et al., 2004) on the VLT on
15-17 Dec. 2016 (program 098.C-0681(C)) and on 10-11 Jan. 2017
(program 098.C-0681(D)), continuing the sequence of {\it Juno}-supporting observations that had started in Feb. 2016 (described in Fletcher et al., 2017b).  The eight filters are selected to provide constraints on upper tropospheric (8--600 mbar) and stratospheric (10--20 mbar) temperatures, along with distributions of 500-mbar aerosols and ammonia gas.  Although these observations were not global in extent, they were designed to capture two separate hemispheres on two separate nights. VLT's 8-m primary mirror provided diffraction-limited spatial resolutions of 0.25-0.8\pr.  Standard image reduction procedures were used (Fletcher et al., 2009), including despiking and destriping to remove detector artefacts, limb fitting to assign geometric information to each pixel, cylindrical reprojection, and absolute radiometric calibration via comparison to Cassini Composite Infrared Spectrometer (CIRS) observations. 
 
\subsection{Subaru}

Images of Jupiter at 7-20 $\mu$m were acquired using the COMICS instrument at the Subaru telescope between January 11 and 14, 2017 (UTC) (Kataza et al., 2000).  Subaru's 8-m primary aperture provides a similar spatial resolution as the VLT at the same wavelengths.  A 2$\times$1 dithering of the COMICS field-of-view ($\sim$45\pr$\times$32\pr) was performed in order to map the entire jovian disk ($\sim$37\pr) while avoiding detector artefacts at the edges of the field.   The reduction of images was performed using the same procedures as described above for the VLT/VISIR data. Images recorded over the four consecutive nights were stitched together to produce an image over 360$^\circ$ in longitude.

\subsection{Gemini}

Thermal infrared images were taken with the NIRI instrument at Gemini North Observatory in the 5-$\mu$m wavelength range (Hodapp et al. 2003). We use the M' filter, with a central wavelength of 4.68 $\mu$m, and the f/32 camera with its 22.4\prs square field of view. Data were acquired on 2017-01-11 (program GN-2016B-FT-18), and are available from the Gemini archive at https://archive.gemini.edu/. 

Images were mapped into the latitude/longitude coordinate space by
aligning the data with a synthetic wire-frame disk, and stacked in the
latitude/longitude coordinate space to avoid errors that would be
introduced by coadding images of a rotating planet. A ``lucky
imaging'' approach was used, taking many 0.3-sec exposures and
coadding only the sharpest individual frames. The full data reduction
pipeline is described by Wong et al. (2019).

\subsection{Keck}

We obtained 5-$\mu$m spectra of Jupiter using NIRSPEC, which is an
echelle spectrograph on the Keck 2 telescope, with 3 orders dispersed
onto a 1024$\times$1024 InSb array (McLean et al 1998). A
0.4\pr$\times$24\prs slit was aligned north-south on Jupiter 
at two longitudes east of the SEB source outbreak, resulting in spectra with a resolving power of 20,000.
The spectra were obtained on January 11, 2017 (program
2016B$_-$N045NS). The geocentric Doppler shift at this longitude was -31.5 km/sec. The water vapor abundance above Mauna Kea was 2.5 precipitable mm along the line of sight to Jupiter, or 1.7 precitable mm in a vertical column. This was derived from fitting telluric lines in both stellar and Jupiter spectra.

\section{Results}\label{results}

\subsection{Longitude-smeared ALMA Maps}\label{smeared1}

\begin{figure*}
\includegraphics[width=32pc]{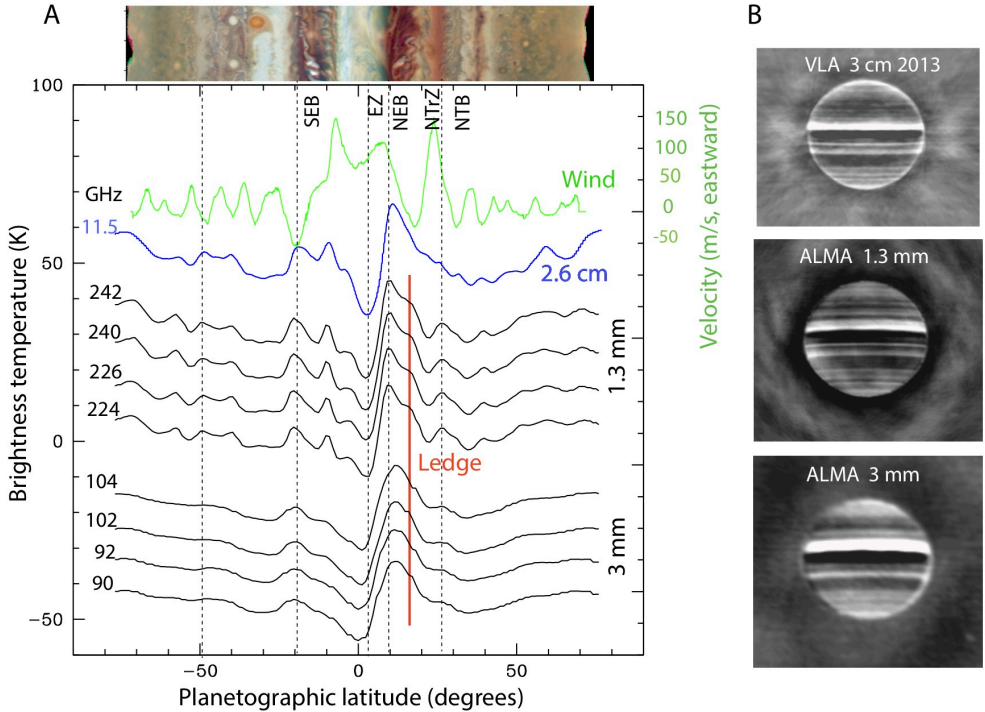}
\caption{A) North-south scans through longitude-smeared ALMA and VLA maps. 
The scans were created by median averaging over 60$^\circ$ of
longitude, centered on the central meridian of each map, after
reprojection on a longitude/latitude grid. Since a limb-darkened disk
had been subtracted from the data, the background level of each scan
is centered near 0 K, as for the 224 GHz scan. The scans are offset
for clarity by 10 K each, while each set is separated by 20 K. The spatial resolution of the 3-mm maps is about 2.5 times lower than at 1.3 mm and the VLA maps, which lowers the feature contrast. 
The vertical dashed lines (at e.g., the EZ, SEB, NEB and NTB) help guide the eye to line up features. The green line at the top is the (eastward) wind profile as measured from the HST data; the scale is given on the right side.
At the top we show a slice through the HST image from Figure~\ref{fig:resolved}. 
B) Longitude-smeared ALMA maps of Jupiter's thermal emission at 1.3 and 3 mm (averaged over the entire Bands 6 and 3, resp.), and a VLA 3-cm map from dP19, after subtraction of a limb-darkened disk.
}
\label{fig:scans}\end{figure*}

Figure~\ref{fig:scans} shows longitude-smeared maps of Jupiter (panel B), averaged over the entire Band 3 (3 mm) and (separately) Band 6 (1.3 mm).  
Similar to the longitude-smeared VLA maps taken in December 2013 (top
image), we see numerous bright and dark bands across Jupiter's disk,
in particular at 1.3 mm where the spatial resolution in the
north-south direction is similar to that of the 2013 VLA data (Table
3). The radio-hot belt at 8.5--11$^\circ$N latitude, near the
interface between the North Equatorial Belt (NEB; 7--17$^\circ$N) and
the Equatorial Zone (EZ; 7$^\circ$S--7$^\circ$N)  is prominent, as well as the minimum in brightness temperature (T$_b$) near a latitude of 4$^\circ$N, i.e., in the EZ.

We reprojected each 2-GHz wide spectral window map on a
longitude/latitude grid, and constructed north-south scans through
each of the maps, which are shown in panel A of Figure~\ref{fig:scans}, together with a VLA scan from 2013 at 2.6 cm (dP19) which probes similar depths as the ALMA scans (see below).
The background level curves upwards at higher latitudes, because the poles are less limb-darkened than east-west scans along the planet, as shown before from VLA maps (de Pater, 1986;  dP19) and Cassini radiometer data (Moeckel et al., 2019). 
The top-most green curve is the wind profile as measured from the HST data
(11 January 2017), using the methodology of Asay-Davis et al. (2011) and Tollefson et al. (2017).
A strip through the HST map (Section~\ref{resolved1}) is shown at the top of the figure. 

\begin{figure}
\includegraphics[width=16pc]{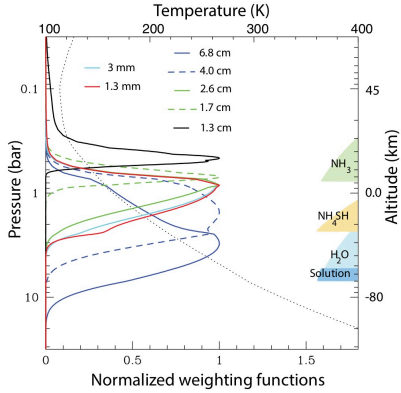}
\caption{The temperature profile (dotted line; wet adiabat) and weighting functions for our nominal atmosphere in thermochemical equilibrium, under nadir viewing conditions. In our nominal atmosphere the abundances of CH$_4$, H$_2$O, and Ar in the deep atmosphere are enhanced by a factor of 4 over the solar values, and NH$_3$ and H$_2$S are enhanced by a factor of 3.2 each. On the right-hand side we indicate the various cloud layers expected to form in thermochemical equilibrium. The weighting functions move down in the atmosphere if there is less NH$_3$ gas, such as in the NEB.
}
\label{fig:contrib}\end{figure}

Ammonia gas is the dominant source of opacity over the entire mm--cm wavelength range, so our maps can be used to derive the 3-dimensional distribution of this gas (as in dP16, dP19). Since
the 1--3 mm and 2.5--3 cm spectral ranges are on opposite sides of the NH$_3$ absorption band, and have a similar absorption strength, the two wavelength ranges probe the same depths in Jupiter's atmosphere ($\sim$0.5--4 bar)
as shown by disk-averaged spectra (Fig. 4 in dP19) and the weighting
functions (Fig.~\ref{fig:contrib}). Despite the 3-year separation
between the 2013 VLA and ALMA data, the similarity between the VLA 2.6 cm and ALMA 1.3 mm scans (at a similar spatial resolution) is striking.
The contrast between the minimum (EZ) and maximum (radio-hot belt, i.e., NEBs) brightness temperatures in the ALMA maps varies from 22 to 27 K from 3 mm down to 1.3 mm (Fig.~\ref{fig:scans}), which is the same as measured with the VLA at 2.5--3 cm (Fig. 7 in dP19). In addition, the zone-belt structure in the southern hemisphere shows an excellent match between the 2013 VLA and ALMA scans. This shows that, averaged over longitude, the southern hemisphere has remained the same over a 3-year period (12/2013--1/2017).

The situation is different in the northern hemisphere. Near
17--18$^\circ$N latitude, there is a ``ledge'' (i.e., a plateau with an abrupt
drop-off) in the ALMA profile
that is missing in the 2013 VLA data. Moreover, the ALMA data show a
clear minimum in T$_b$ at $\sim$23$^\circ$ in the North Tropical Zone
(NTrZ; 17--24$^\circ$N), just south of the prominent eastward jet;
such a clear minimum is not seen in the 2013 VLA data. In the visible,
the NTrZ is usually white, indicative of upwelling gases. At present,
as shown in the HST strip above the scans, the NTrZ is highly
disturbed (and in part colored orange). A comparison between the ALMA scans and the HST strip further shows that subtle variations in T$_b$ match variations in color in the HST data, which implies that changes in the visible are related to latitudinal variations in the ammonia abundance below the cloud deck. This is also interesting, because it is well known that colors of Jupiter's bands change temporally (e.g., fading of the SEB, expansions of the NEB, and the color change in the NTrZ in Fig.~\ref{fig:scans}A), while the wind profile (at the NH$_3$ cloud deck) is very stable (except for changes in the absolute velocity of the 24$^\circ$ eastward jet) (e.g., Rogers, 1995; Asay-Davis et al., 2001; Tollefson et al., 2017).

\subsection{Longitude-resolved Maps at Radio, Visible and Mid-Infrared
Wavelengths}\label{resolved1}

\begin{figure*}
\includegraphics[width=32pc]{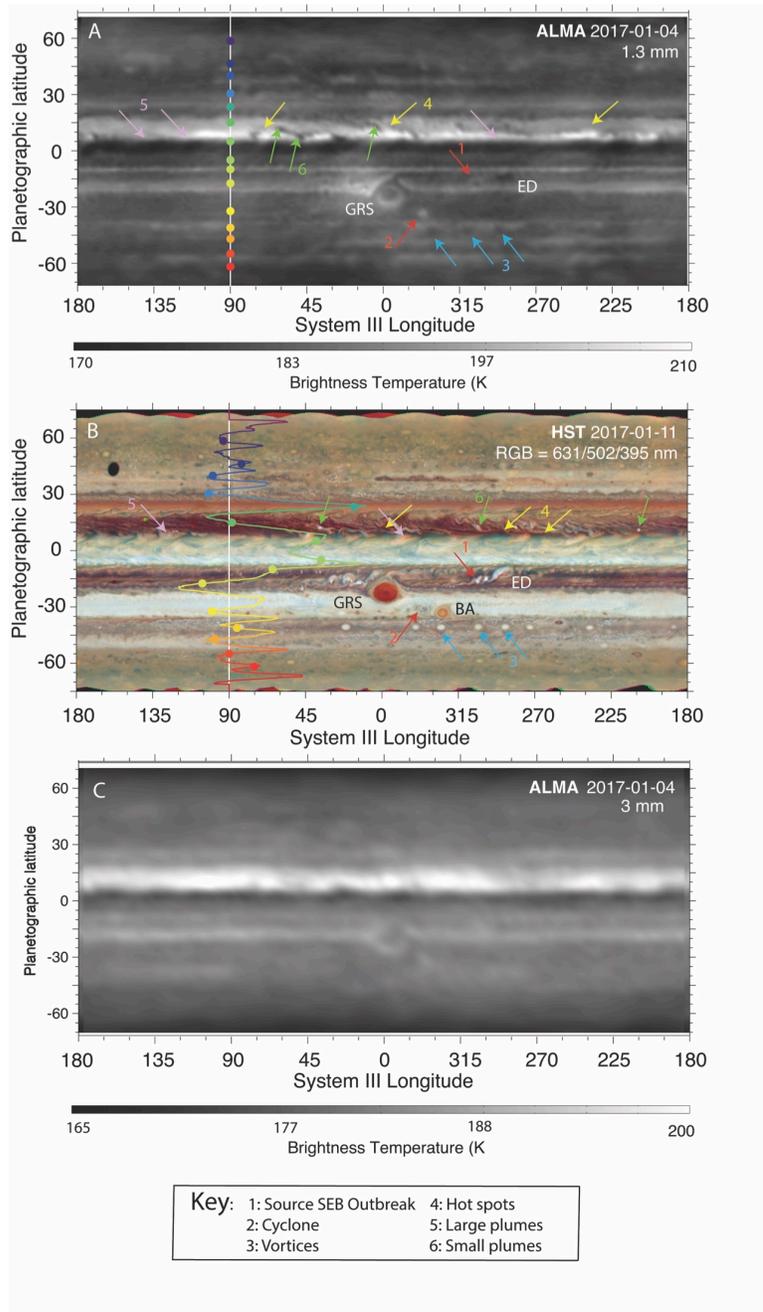}
\caption{a) ALMA map at 1.3 mm, constructed from data taken 3-5 Jan. 2017.
b) HST map from 11 Jan. 2017, with the zonal wind profile superimposed.  c) ALMA map at 3 mm, constructed from data taken 3-5 Jan. 2017.
Various features are indicated, such as the iconic GRS and Oval BA. Features 1--6, indicated by different colored arrows with numbers, are indicated in the Key. Since features move at different speeds across Jupiter's disk, and the ALMA and HST observations were taken on different days, we indicate on the HST panel how features at different latitudes (colored dots) have moved since the ALMA data were taken. 
}
\label{fig:resolved}
\end{figure*}

\begin{figure*}
\includegraphics[width=32pc]{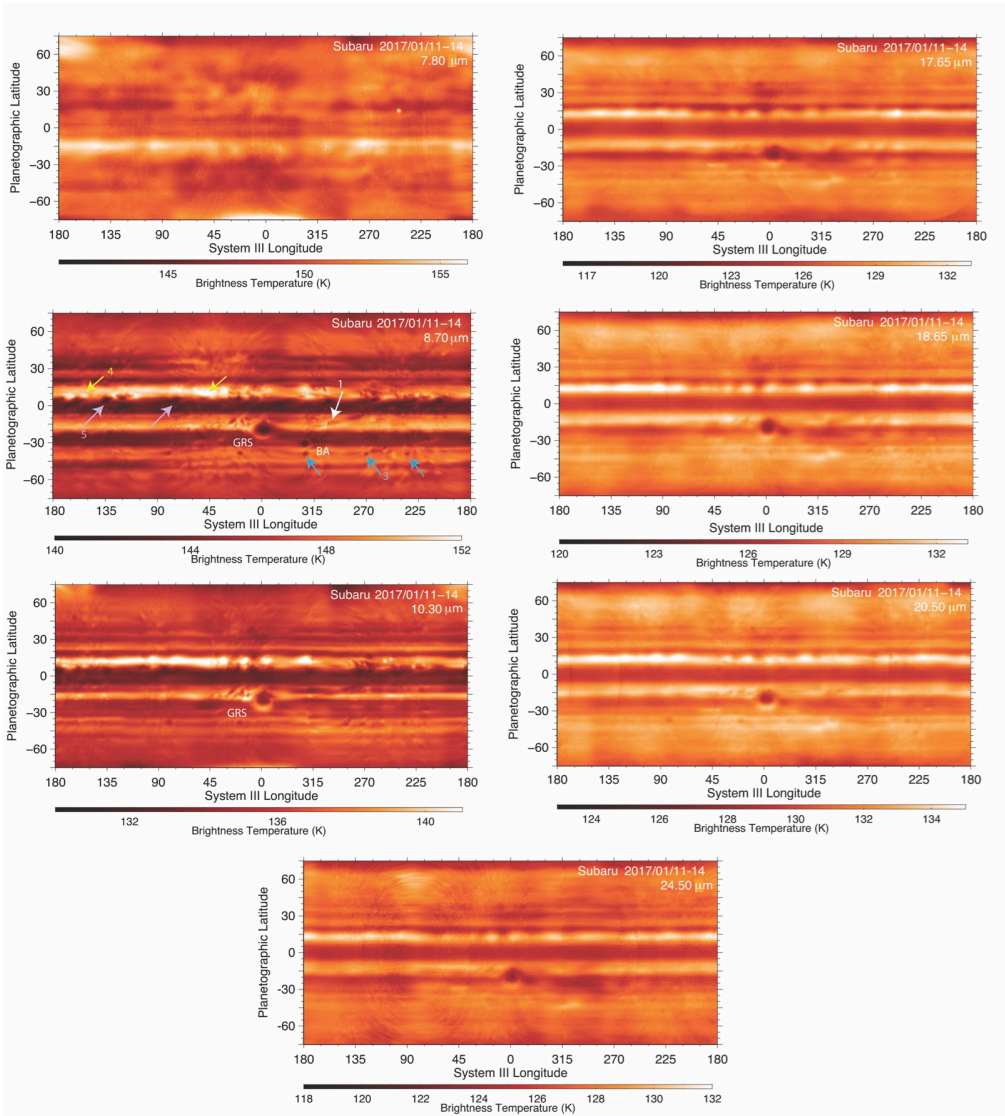}
\caption{Subaru/COMICS maps at wavelengths of 7.8--24 $\mu$m. Several
  features are indicated on the 8.70-$\mu$m map, with the same Key as
  in Fig.~\ref{fig:resolved}.
}
\label{fig:comics}
\end{figure*}

\begin{figure}
\includegraphics[width=16pc]{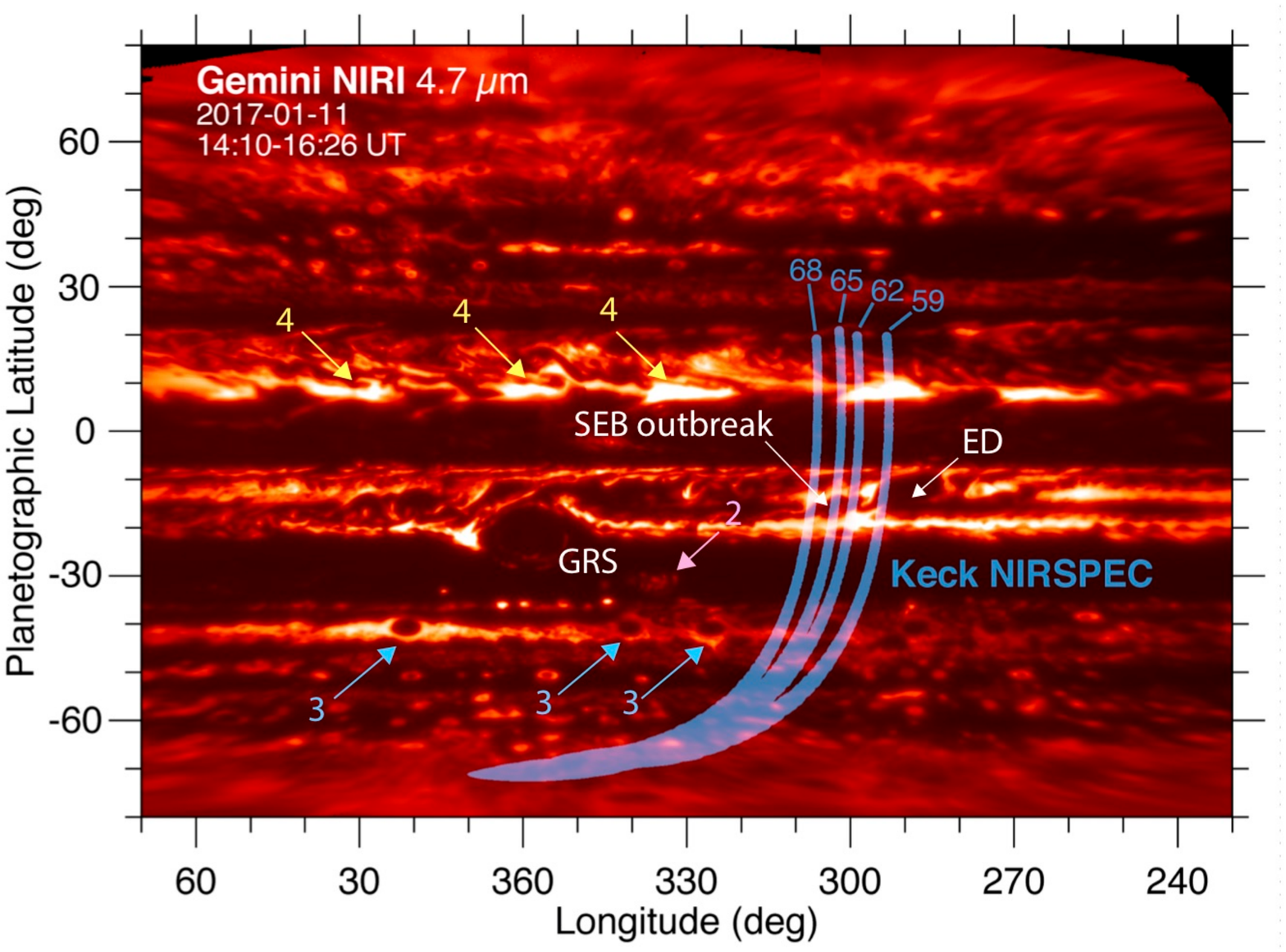}
\caption{A 4.7-$\mu$m map of Jupiter constructed from images taken by
  Gemini/NIRI on 11 Jan. 2017.  The NIRSPEC slit positions are
  projected onto the map; the data at the two telescopes were taken
  simultaneously. We analyzed the ground tracks denoted 59 and 65 in
  Section \ref{sec:rtkeck}. The numbers with arrows refer to the same Key as in Fig.~\ref{fig:resolved}.
}
\label{fig:niri}
\end{figure}

A plethora of structure is seen in the ALMA disk-resolved maps, as shown most clearly in the 1.3-mm maps, Figure~\ref{fig:resolved}A. 
Bright areas indicate a higher brightness temperature, assumed to be
caused by a lower NH$_3$ abundance (as in dP19, assuming the
temperature profile follows an adiabat), and dark areas indicate a
lower brightness temperature, caused by a higher opacity in the
atmosphere. The radio-hot belt at 8.5--11$^{\circ}$N latitude (NEBs)
contains prominent hot spots with small well-defined dark regions interspersed. The dark regions are small plumes of NH$_3$ gas, which are likely associated with the small bright clouds in the HST map (green arrows, \#6, in Fig.~\ref{fig:resolved}B) at similar latitudes ($\sim$12$^\circ$N). 
Just to the south are larger dark and somewhat oval-shaped regions;
these are the plumes of ammonia gas that were most striking in VLA
data at $\sim$6 cm wavelength (dP16) but were visible at all radio
wavelengths observed (1--13 cm) (dP19), as well as in the thermal
infrared as indicated in Figure~\ref{fig:comics}, in particular near
10 $\mu$m (Fletcher et al., 2016). 
The Great Red Spot (GRS) in the ALMA map is a well-defined structure
surrounded by a bright ring, and a turbulent wake to the west. Oval BA
is not visible, likely due to the absence of the, apparently
transient, bright ring around the feature and westward wake, which
made it visible in the 2013-2014 VLA data (dP16, dP19). A small
cyclonic vortex can be discerned to the west of Oval BA, that is
bright at radio wavelengths and at 5 $\mu$m (see Fig.~\ref{fig:niri}), indicative of NH$_3$-dry air and a clearing of
aerosols. The anticyclonic vortices at 40$^\circ$S are characterized
in the radio and mid-infrared by a darker center surrounded by
brighter areas. The dynamics of these small vortices as seen at 5
$\mu$m was discussed by de Pater et al. (2010; 2011). The HST and mid-infrared data were taken $\sim$1 week after the ALMA observations. The colored line on the HST panel traces Jupiter's wind profile, and hence aids both in identifying how much features have moved, and to distinguish latitudes of cyclonic from anticyclonic wind shear.

\section{Radiative Transfer (RT) Models}

In the following we discuss radiative transfer (RT) model results of
our different observations. We start with general model results for
the ALMA data, including the SEB outbreak, and then present more
specific calculations at visible and 5 $\mu$m wavelengths with regard
to the SEB outbreak.

\subsection{RT Modeling of the ALMA Longitude-smeared Maps}\label{sec:rt_sm}

We model our data with the RT code 
Radio-BEAR (Radio-BErkeley Atmospheric Radiative transfer)\footnote{https://github.com/david-deboer/radiobear}, described
in detail in de Pater et al. (2005; 2014; 2019). As in dP19, in our
nominal atmosphere, assumed to be in thermochemical equilibrium, the
abundances of CH$_4$, H$_2$O, and Ar in the deep atmosphere are
enhanced by a factor of 4 over the solar values, and NH$_3$ and H$_2$S
are enhanced by a factor of 3.2, and the temperature-pressure (TP)
profile follows an adiabat (typically wet in zones, dry in belts),
constrained to be 165 K at the
1 bar level to match the Voyager radio occultation profile (Lindal,
1992). At  pressures $\lesssim$0.7 bar, the TP profile follows that determined from mid-infrared ({\it Cassini}/CIRS) observations (Fletcher et al., 2009).

As discussed in dP19, variations in the observed brightness
temperature can in principle be caused by
variations in opacity or by spatial variations in the physical
temperature. They show that variations in opacity are much more likely
than changes in temperature, and therefore, like in dP19, we attribute
all changes to variations in opacity. The latter authors also
investigated the effect on the brightness temperature due to changes
in the TP profile at and above the ammonia clouddeck, as sensed at
mid-infrared wavelengths. After changing the
TP profile at each latitude to that observed by {\it Cassini}/CIRS
(Fletcher et al., 2016), they showed only small
changes (varying from zero to perhaps up to maximal 5 K in brightness
temperature at some latitudes) near the center of the ammonia
absorption band, between 18 and 26 GHz ($\sim$1.3 cm).
At deeper levels below the NH$_3$ cloud, an equatorial 
thermal wind analysis constrained by Galileo Probe 
vertical wind shear (Atkinson et al. 1998, Marcus et 
al. 2019) suggested that there may be horizontal 
temperature variations of $<$ 3 K between the equator 
and 7.5 N. Our analysis of ALMA data did not consider 
small horizontal temperature differences of this 
magnitude, particularly since vertical wind shear 
cannot be measured in the region of the SEB outbreak.

\begin{figure*}
\includegraphics[width=32pc]{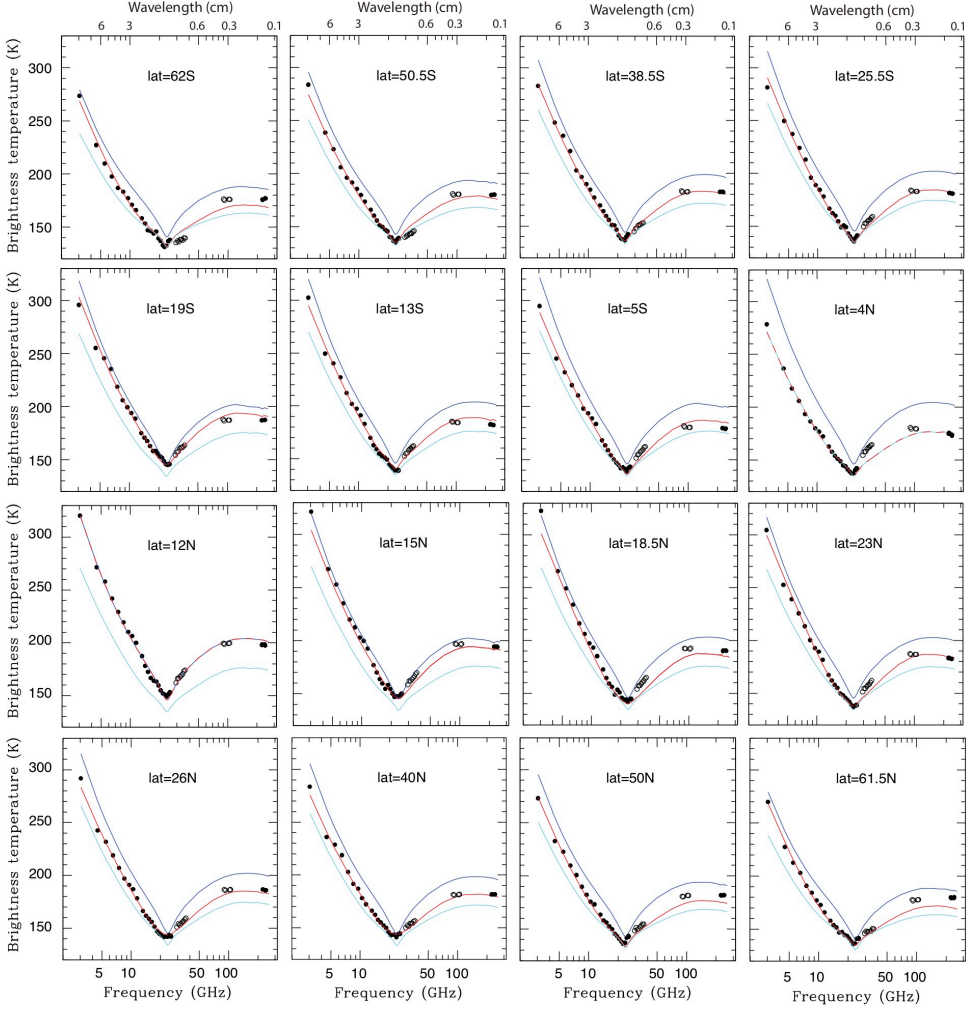}
\caption{Longitude-averaged brightness temperatures T$_b$ (black dots)
  from the VLA 2013-2014 data (3--37 GHz, 10--0.8 cm) and the ALMA
  data (90--242 GHz; 3--1.3 mm) with superposed the best-fit model
  spectra as derived from the VLA data (dP19: red lines) for several
  latitudes. The data at $\sim$30--100 GHz (1--0.3 cm; open circles)
  have a roughly 3--4 times lower spatial resolution than at other
  frequencies. The cyan and
  blue curves on all panels show the models with the parameters
  (NH$_3$ profiles) that gave best fits to the 2013-2014 VLA EZ and
  NEB data. (the red curve coincides with the cyan at 4.0N, and with
  the blue one at 12.0N latitude).
  The spread between data points is a good estimate for uncertainties
  in the data.
}
\label{fig:modelfits}\end{figure*}

\begin{figure*}
\includegraphics[width=32pc]{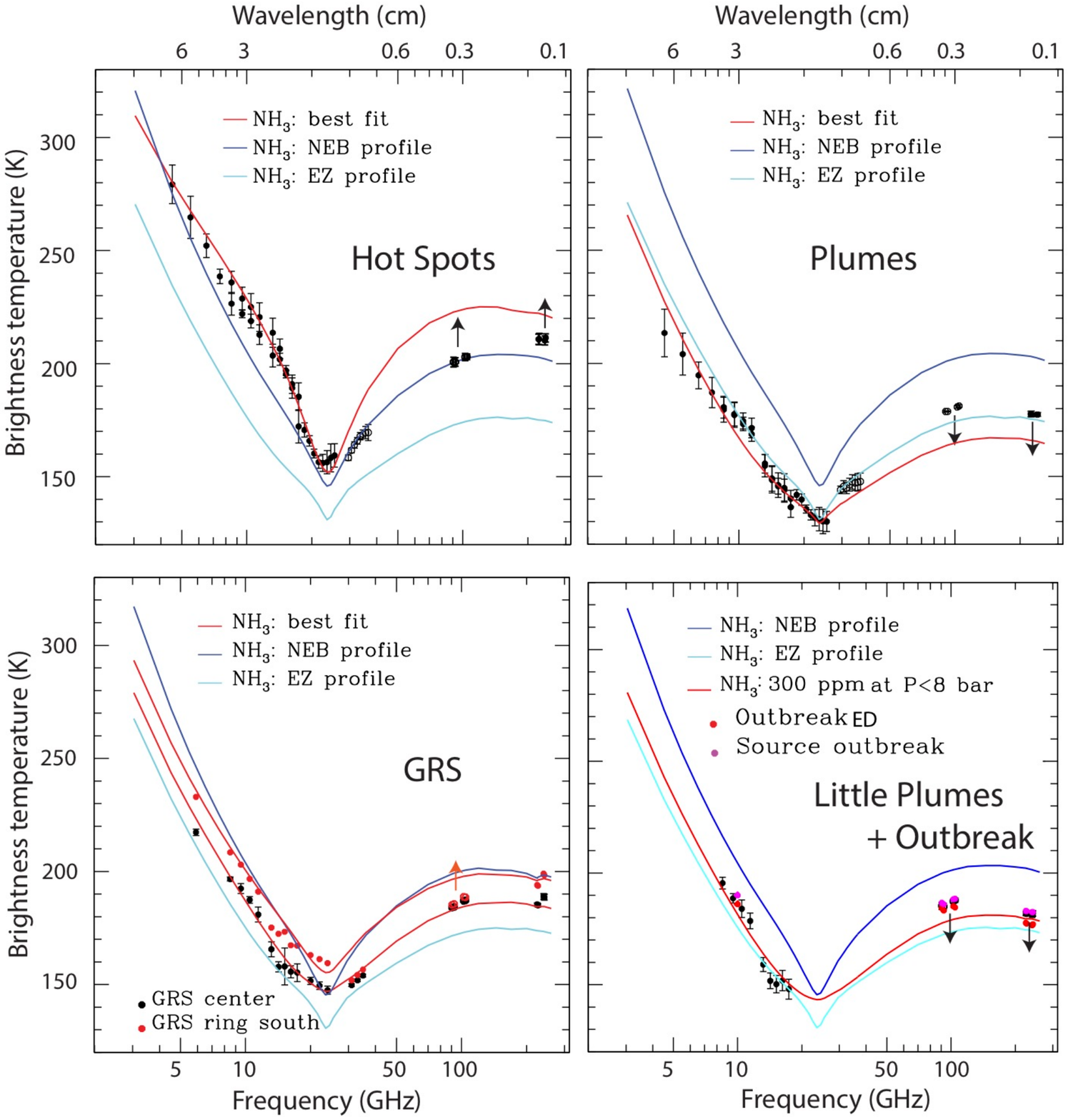}
\caption{Spectra are shown for various features: hot spots, large
  plumes, the GRS (center and ring south of the GRS), little plumes
  and the SEB outbreak, with superposed models that gave best fits to
  the VLA data of the hot spots, GRS, and plumes (red curves). The
  ammonia abundances for all models were shown in dP19. The cyan and
  blue curves on all panels show the models with the parameters
  (NH$_3$ profiles) that gave best fits to the 2013-2014 VLA EZ and
  NEB data. The red curve for the SEB Outbreak data is one where the NH$_3$ abundance at $P < 8$ bar has been replaced by 300 ppm, and following the saturated vapor curve (with a 1\% relative humidity) where appropriate. 
}
\label{fig:modelfits2}\end{figure*}

To examine the 3-dimensional distribution of ammonia gas, or more
specifically to identify changes in this distribution since December
2013, we compare in Figure~\ref{fig:modelfits} the brightness
temperature of the 1--3 mm ALMA data with best-fit models to the
2013-2014 VLA data (from dP19). We stress here that no new models were
produced; the existing models were merely extended into the
mm-wavelength range. Hence, as in dP19, we ignored opacity by clouds. The
latter authors justified this assumption based upon disk-averaged
spectra at mm--cm wavelengths. They argued that, if cloud opacity is important, the brightness
temperatures at mm wavelengths should be affected much more than in
the cm range, since the mass absorption coefficient is inversely
proportional to wavelength for particles that are small compared to the
wavelength (Gibson et al., 2005). 

Figure~\ref{fig:modelfits} shows the zonal-mean brightness temperature
spectra of the ALMA data together with the corresponding 2013-2014 VLA
data, superposed on the models that gave a best fit to the 2013-2014
VLA data at the different latitudes. For comparison we show in all
plots the best fits to the EZ (cyan) and NEB (radio-hot belt; blue),
while the best-fit VLA models are shown in red. The 3-mm data, with a
2.5--3 times lower spatial resolution, show lower limits to brightness
temperatures where maxima in T$_b$ are measured, and upper limits
where T$_b$ minima are recorded. As shown, the ALMA data show a near-perfect match to the red curves,
except perhaps at the highest latitudes. The
brightness temperatures at these high latitudes might be
slightly too high due to the bowl-like structure under the planet
as introduced by missing short spacings (e.g., de Pater et al., 2001;
dP19).

We note that in particular in the EZ (4$^\circ$N), NTrZ
(23$^\circ$N), and at latitudes 30--40$^\circ$N and S  the ALMA data match the VLA
models perfectly, which would corroborate dP19's
assumption that clouds do not affect Jupiter's brightness temperature
at mm-cm wavelengths. To check this statement, we performed several 
RT calculations. These show that in the NH$_3$-rich EZ, contribution functions
peak at such high altitudes that clouds do not affect the modeled
brightness temperature at mm wavelengths. In the NEB, mm-wavelength
observations can penetrate to the level of the NH$_4$SH cloud. We tested one case with high NH$_4$SH mass loading ($\sim$1.6 
g cm$^{-2}$ between 2.4 and 0.9 bar)  (the
water cloud has no effect at mm wavelengths). Although NH$_4$SH cloud 
opacity lowered brightness temperatures by a few 
degrees at mm wavelengths\footnote{As pointed out by de Pater and
  Mitchell (1993), not much is known about the complex index of refraction ($m$) of the
cloud layers. In our 
calculations we used $m = 1.7 -0.005j$ for NH$_4$SH. De Pater and Mitchell (1993)
show results for $m = 1.7 - 0.05j$.}, an extremely strong updraft 
(length scale $\sim$30 km; see Wong et al., 2015) would be required to generate 
this much cloud mass. The low NH$_3$ abundance in the NEB, down to
over the 20 bar level (dP16; dP19;
Li et al., 2017), is suggestive of subsiding rather than rising air,
which makes the presence of such a thick NH$_4$SH cloud layer quite unlikely. 
We therefore interpret deviations in the 
ALMA data compared to the model spectrum in terms of variations in the
NH$_3$ abundance, and ignore potential effects of cloud opacity.

At latitudes 5--13$^\circ$S the ALMA brightness temperatures are
slightly lower, so there
may have been slightly more NH$_3$ gas below the cloud layers than in
2013-2014. At 12$^\circ$N the ALMA brightness temperature is a
tad too cold, and at 18$^\circ$N and 26$^\circ$N it is slightly 
warmer than the models, i.e., there seems to be less NH$_3$ gas
at 18$^\circ$N and 26$^\circ$N in 2017 then in 2013-2014. This would
explain the observation that the minimum in T$_b$ at 23$^\circ$N
appears to be more pronounced in the ALMA data then in the 2013 VLA data (Fig.~\ref{fig:scans}A), since the NH$_3$ at latitudes north and south of 23$^\circ$N have changed, while the NH$_3$ abundance stayed constant at 23$^\circ$N. 

\subsection{RT Modeling of the ALMA Longitude-resolved Maps}\label{sec:rt_res}

Figure~\ref{fig:modelfits2} compares several resolved features in the
ALMA data to models that best fit the 2013-2014 VLA data (from dP19)
for the same type of features; as for Figure~\ref{fig:modelfits}, these
models were obtained with our RT code Radio-BEAR. The red curves show
the best fit models to the VLA data, and the cyan and blue curves show
the best fits to the longitude-smeared EZ and NEB, respectively. In order to properly compare
the ALMA data to the models, however, we need to take into account the spatial
resolution of the data. For the VLA 2--4 cm data, this varied roughly
from 1000$\times$1000 to 2000$\times$2000 km$^2$, while the
resolution of the 1.3-mm ALMA data is $\sim$2000$\times$4000 km$^2$,
and for the 3-mm data it is 2.5 times lower still
($\sim$5000$\times$8600 km$^2$) (Table 4).
As shown, the model for the center part of the
GRS, which is quite extended, fits the ALMA data very well, and the
ALMA 1.3 mm data for the bright ring on the south side of the GRS
also agree well with the model. As expected, the brightness
temperatures of the latter at 3 mm, like at
0.9 cm (30-35 GHz), are too cold compared to the models because the
ring is not resolved in these observations. Similarly, the Hot Spots
indicate too low a T$_b$ at both 1.3 and 3 mm, and too high a T$_b$
for the plumes. We also indicate the T$_b$ for the source of the SEB
outbreak and the disturbance to the east of the outbreak, referred to
as the ED (East Disturbance). Because these features are small in
angular extent, the measured brightness temperatures should be
considered upper limits. We also indicate the values from the January
2017 VLA data (discussed further in Section 5.2), which are
well-aligned with the values for the little plumes as measured in
the 2013-2014 VLA data. The red curve on this graph is not a best fit;
instead, it is a model where the NH$_3$ abundance was assumed to be
300 ppm at pressures $P <$ 8 bar. 

\begin{figure*}
\includegraphics[scale=0.9]{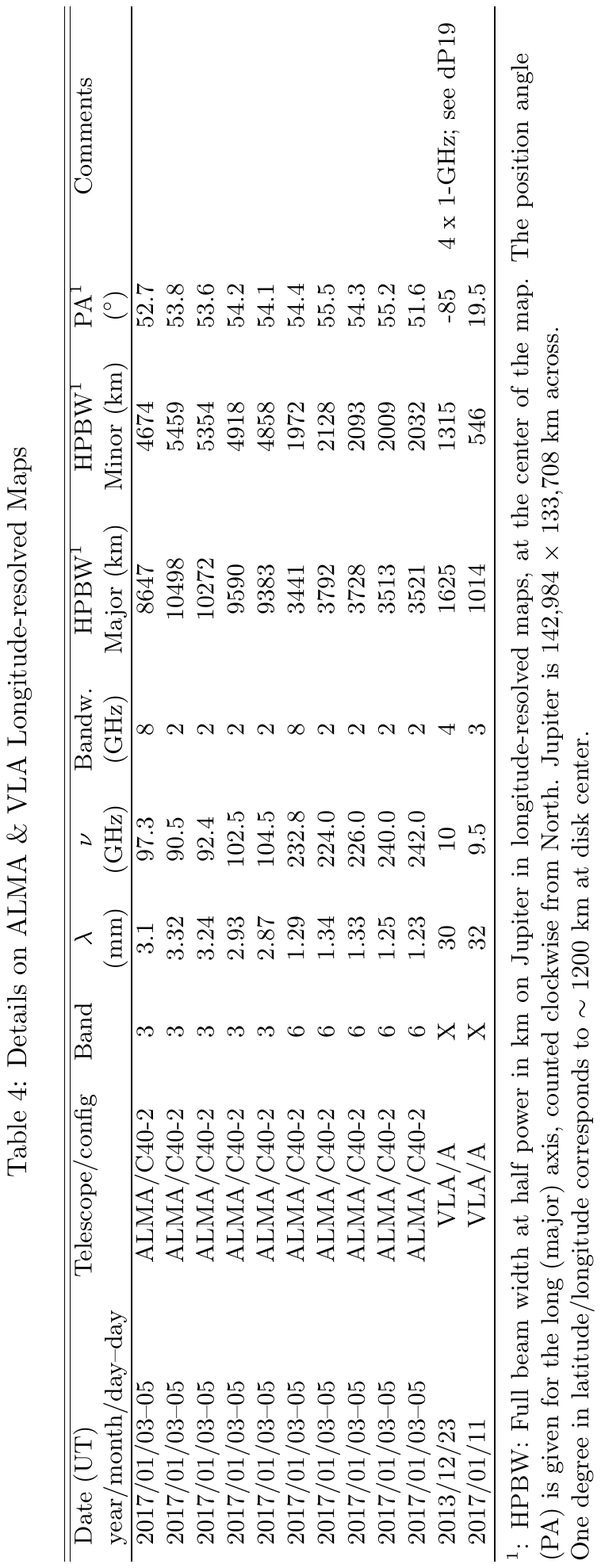}
\end{figure*}

\subsection{SEB Outbreak in HST Data}\label{sec:rt_res2}

We chose the brightest spot at visible wavelengths as the location of the outbreak (see Fig.~\ref{fig:resolved}).  The outbreak spectrum was constructed by taking the I/F value at said location in each filter.  Background spectra represent the average value of three different locations close to the outbreak. These spectra were fit using our in-house RT code
SUNBEAR (Spectra from Ultraviolet to
Near-infrared with the BErkeley Atmospheric Retrieval)
(Luszcz-Cook, et al., 2016), a \lstinline{python} program based on the
\lstinline{pydisort} module \footnote{https://github.com/adamkovics/atmosphere/blob/master/atmosphere/rt/pydisort.py} (\'Ad\'amkovics et al., 2016).
SUNBEAR has been used to model Uranus at IR wavelengths
(de Kleer, et al., 2015) and Neptune at UV, Visible, and IR wavelengths
(Luszcz-Cook, et al., 2016; Molter, et al., 2019).
SUNBEAR takes as inputs the temperature-pressure profile, atmospheric
composition as a function of depth, a model of the aerosols, and the
gas opacities as a function of temperature and pressure.
These inputs are used to construct a model atmosphere, which is fed
into \lstinline{pydisort} to solve the radiative-transfer equation.
Further details on the code can be found in Appendix A of
Luszcz-Cook, et al., (2016). 

Both the background and outbreak models consist of a variable number
of haze layers, an NH$_{3}$-ice cloud, and an NH$_{4}$SH cloud.
The scattering properties of all haze layers were derived from Mie
theory, with just the particle size and imaginary
refractive index as variable inputs.
The fraction of particles with radius $R$, given a peak particle size
$r_{p}$, is given by
\begin{equation}
  \label{eq:size_dist}
  p(R) = \frac{R^{6} e^{\frac{-6R}{r_{p}}}}{\sum_{r=0}^{r=\infty}r^{6} e^{\frac{-6r}{r_{p}}}}
\end{equation}
The real part of the refractive index was set to that of ammonia ice, i.e., $1.4$.
The clouds were modeled as perfect reflectors with the Henyey-Greenstein
asymmetry parameter $g=-0.3$.
The NH$_{3}$-ice cloud was placed at $0.7$ bar and given an opacity of
$10$, while the NH$_{4}$SH-ice cloud was placed at $2.5$ bar and given
an opacity of $30$. We added four haze layers above the clouds. The topmost haze layer extended from 1 to 100 mbar, the second from 100 to 200 mbar, the third from 200 to 650 mbar, and the fourth from 650 mbar to 700 mbar. We adapted the opacities and particle radii of these haze layers to fit he spectra.

To fit the background spectra, we used an
imaginary index of refraction similar to that used for the NEB in Fig. 7 of
Irwin et al., (2018).
The peak particle radii for the haze layers, which we will refer to as
hazes 1--4, with 1 being the uppermost and 4 being the lowermost,
were $r_1= $0.1 $\mu$m, $r_2= $0.3 $\mu$m, $r_3= $0.8 $\mu$m, and $r_4= $1.0 $\mu$m,
The cumulative opacities for each layer were  $\tau_{1} = 0.1$,
$\tau_{2} = 0.1$,  $\tau_{3} = 1.8$, and  $\tau_{4} = 3.7$.

\begin{figure*}
\includegraphics[width=32pc]{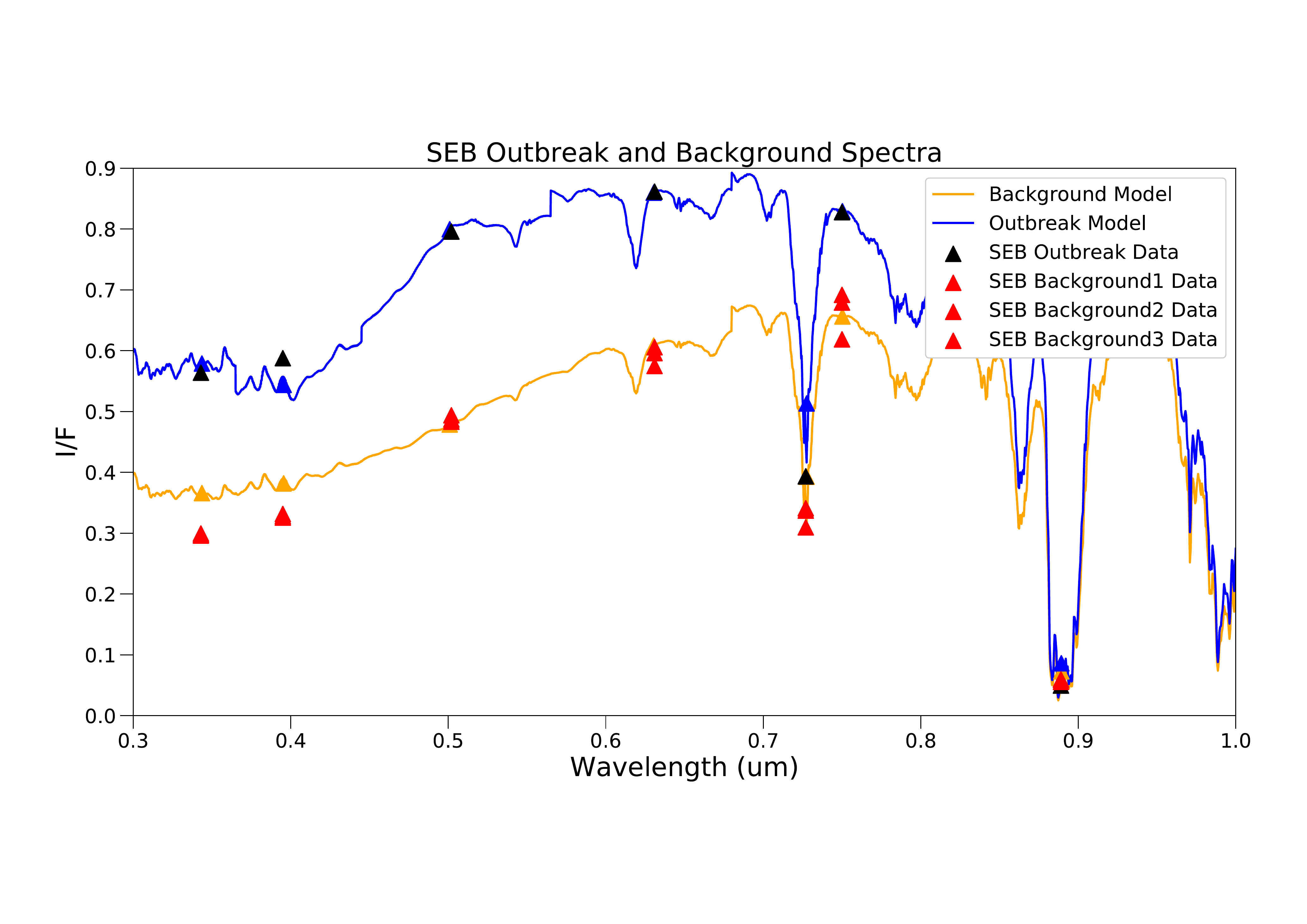}
\caption{SUNBEAR RT model fit to the SEB outbreak plume and background. The
    outbreak and background data are shown in black and red, and the
    respective models are shown in blue and orange. The blue and orange
    triangles show the model values expected in the HST filters (i.e.,
    model convolved with the HST filter shape).
}
\label{fig:hstmodel}
\end{figure*}

To match the outbreak spectrum, we used an imaginary refractive index
of $4\times 10^{-3}$ in the UV,
$7.5\times 10^{-4}$ in the visible, and $2 \times 10^{-3}$ in the IR.
Using the same haze labeling as the background model, the peak
particle radii are $r_1= $0.1 $\mu$m, $r_2= $0.8 $\mu$m, $r_3= $0.3
$\mu$m, and $r_4= $0.8 $\mu$m.
The cumulative opacities are $\tau_{1} = 0.1$,
$\tau_{2} = 0.2$,  $\tau_{3} = 0.62$, and  $\tau_{4} = 0.85$.
The results of the radiative transfer modeling of these atmospheres
are shown in Figure~\ref{fig:hstmodel}.

We find that the haze above the SEB outbreak has twice the cumulative opacity of the
background model at high altitudes (100--200 mbar), but barely a
quarter of the cumulative opacity at 0.2--0.7 bar.
The outbreak plume also has
different scattering properties from the background atmosphere.
The imaginary index of refraction is higher for the outbreak, indicating that the hazes are more prone to absorbing light
at these wavelengths than the background atmosphere. Finally, we see variations in peak particle radius for each layer that suggests larger particles are being transported to the tropopause from deeper down in the atmosphere, while simultaneously removing larger particles from these deeper hazes.

\begin{figure*}
\includegraphics[width=32pc]{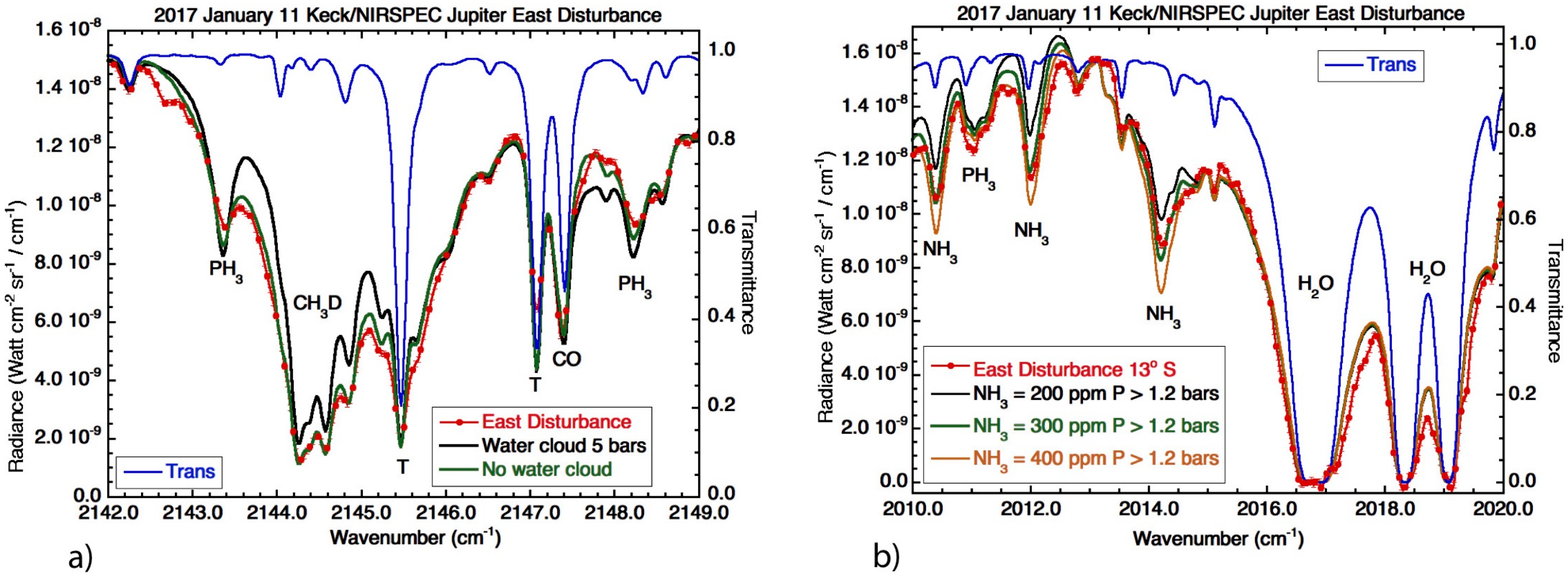}
\caption{a) Spectrum of Jupiter's ED compared with synthetic spectra
  generated from models with (black) and without (green) opaque clouds
G at 5 bars. The best fit to the CH$_3$D feature is for the model
  without a cloud at the level where water is expected to condense on
  Jupiter.
  b) Spectrum of the ED compared with synthetic spectra generated from models with three abundances (200, 300, and 400 ppm) of gaseous NH$_3$. The best fit model has 300 ppm NH$_3$. All models have an H$_2$O abundance of 47 ppm for P $>$ 4.5 bars, consistent with that measured in the Galileo Probe entry site, and no opaque cloud at the 5-bar level.
}
\label{fig:keck}
\end{figure*}

\subsection{Keck 5-$\mu$m Spectroscopy of the SEB Outbreak}\label{sec:rtkeck}

In Figure~\ref{fig:niri}, we project the ground tracks of the NIRSPEC
slit onto a 4.7-$\mu$m Gemini/NIRI image of Jupiter taken at the same
time. We analyzed the tracks denoted 59 and 65, the file numbers of the NIRSPEC
spectra. Note that track 59 traverses the western portion of the
East Disturbance (ED) at 13$^\circ$S, and track 65 traverses the
source region of the SEB outbreak. Both areas are dark at this wavelength due to higher cloud opacity. 
We analyzed all 3 NIRSPEC orders centered on 4.66, 4.97, and 5.32
$\mu$m. The 4.66-$\mu$m spectrum reveals spectrally resolved
absorption features of CH$_3$D, which were used to derive cloud
structure. Spectra at 4.97 $\mu$m show gaseous H$_2$O and NH$_3$
absorption features formed between 4 and 6 bars on Jupiter. The
5.32-$\mu$m order samples a strong NH$_3$ absorption band permitting
retrieval of the albedo of the upper cloud layer.
In the following we discuss our model fits to the spectra; the details
of our methodology are described by Bjoraker et al. (2018).

Figure~\ref{fig:keck}a shows a portion of the spectrum of the ED at 4.66 $\mu$m (2142 to 2149 cm-1). We compare the observed spectrum with a model containing an opaque water cloud at 5 bars, and an alternate model with no cloud opacity at this level. Both models included sunlight reflected from an upper cloud at 600 mbars with an albedo of 12\%. This albedo was obtained by fitting the spectrum of the ED in a strong NH$_3$ band at 5.32 $\mu$m (not shown) where the radiance from the thermal component is expected to be zero. The observed CH$_3$D absorption feature at 2144 cm$^{-1}$ is broader than that in the opaque water cloud model, while the observed line shape is fit quite well by the model lacking a water cloud.

Using a cloud model with a reflecting layer at 600 mbars that is
partially transmitting to allow thermal radiation from the deep
atmosphere to emerge, we next investigated the abundance of gaseous
H$_2$O and NH$_3$ by fitting the NIRSPEC spectrum at 4.97 $\mu$m
(2010-2020 cm$^{-1}$). Since we found no evidence of a water cloud at
5 bars, we adopted the gaseous H$_2$O profile as measured in the
Galileo Probe entry site (Wong et al., 2004). We adjusted one
parameter, namely the pressure above which the H$_2$O abundance is
equal to zero. The best fit was for a pressure of 4.5 bars. At deeper
levels we adopted the Galileo Probe mole fraction of 47 ppm
H$_2$O. Once we obtained a good fit to the wing of the strong H$_2$O
absorption line near 2016 cm-1, we iterated on the deep mole fraction
of NH$_3$. In Figure~\ref{fig:keck}b we compare the observed spectrum
of the ED to synthetic spectra calculated from models with 200, 300,
and 400 ppm NH$_3$. The best fit was for 300 ppm NH$_3$ for pressures
greater than 1.2 bars.

The spectra for track 65 were essentially the same as for 59, and
hence the same results were obtained; i.e., all our spectra are well-matched with a model with thick clouds
at $\sim$600-mbar level, no cloud near 5-bar, and a NH$_3$ abundance
of $\sim$300 ppm. However, at this point we should consider possible contamination by nearby
hot regions, since emissions from higher-temperature regions would
dominate the intensity at this wavelength (e.g., Wien's law). Indeed, 
normalized 5-$\mu$m spectra of nearby Hot Spots (not shown) are nearly
identical to those of the ED and the source of the SEB outbreak.
We can evaluate our 5-$\mu$m fluxes by comparing the ratio of flux at
4.7 $\mu$m between Hot Spots and the ED in both the Gemini/NIRI images
and in the NIRSPEC spectra. A Hot Spot at 17$^\circ$ S, 294$^\circ$W
is 12 times brighter than the ED in the NIRI image. The corresponding
ratio in the 4.7-$\mu$m continuum level in the NIRSPEC data is about
6. The integration time for the NIRSPEC spectrum was 30 seconds,
vs. the much shorter time (0.3 second) in the NIRI image. We also
observed some westward motion of the slit by comparing images taken
before and after the spectral integration. Moreover, the spectra were
taken using conventional spectroscopic techniques, while the NIRI
image shows a much higher spatial resolution (essentially diffraction
limited). This effects our
interpretation of both gas abundances (we should consider our values
as lower limits) and the absence of a deep cloud (i.e., there may well
be a deep cloud).

\section{Discussion}

It has been well established that the belts in Jupiter's atmosphere are regions of episodic violent convective eruptions, sometimes associated with lightning events (e.g., Vasavada and Showman, 2005; Brown et al., 2018). The eruptions show up as bright plumes at visible wavelengths.
Such vigorous eruptions require the existence of a large reservoir of
convective available potential energy (CAPE; Emanuel, 1994), that can
be released through moist convection. CAPE is produced by radiative
cooling in the upper atmosphere ($P \lesssim$1 bar) over a radiative
timescale (4--5 years on Jupiter; Conrath et al., 1990). Showman and
de Pater (2005) discuss that in the belts, regions that are dominated
by subsiding dry air, the virtual potential temperature (i.e., the
temperature dry air would have if its pressure and density were equal
to that of moist air) may slightly exceed that of the deep (dry)
adiabat with an interface below the water cloud. This slight jump in
potential temperature (i.e., mainly caused by the change in mean
  molecular weight due to condensation of water) forms a stable layer that inhibits vertical mixing there. Occasionally, plumes may rise up to the (water) condensation level, where latent heat produced upon condensation may propel the plumes further up along a moist adiabat, thereby reducing CAPE. Due to the presence of the stable layer below the water condensation level, CAPE cannot be completely depleted, and an equilibrium is set up between the rates at which CAPE is produced and dissipated. This has been modeled numerically by Sugiyama et al. (2014).

\subsection{Moist Convection in the NTB, and the NEB Expansion}

In October 2016 four super-bright plumes were spotted  on the south side of the NTB, moving with the fast 24$^\circ$N eastward jet. These plumes signified the onset of a large disturbance, or reorganization, of the NTB, as recorded subsequently by the amateur-astronomy community, leading ultimately to the orange-colored band seen in the HST map (Fig.~\ref{fig:resolved}B) (see S\'anchez-Lavega et al., 2017 for a full description and numerical simulation of events). Such disturbances have been recorded in the NTB roughly every 5 years (Rogers, 1995; Fletcher, 2017), i.e., consistent with the build-up of CAPE. To this date, we have {\it no} observations that trace the plumes down to below the cloud layers, however.

As mentioned in Section~\ref{sec:rt_sm}, the NH$_3$ abundance had not
changed in the NTrZ between Dec. 2013 and Jan. 2017, but it had slightly
decreased at 18.5$^\circ$N (the ledge in the NEB) and in the NTB. It
may be possible that the super-bright plumes were rising up so fast
that condensation did not start until well above the ammonia cloud
deck. As shown in Section~\ref{sec:rt_res}, plumes indeed rise up well
above the ammonia cloud layer. With the low temperatures at these high altitudes, the air would become very dry upon condensation. This very dry air could descend in the neighboring belt regions (NTB, NEB-ledge), causing them to be dryer than under normal circumstances. 

Fletcher et al. (2017b) showed that in 2015-2016
the brown color of the NEB expanded northwards (from 17$^\circ$ to
20$^\circ$),  into the NTrZ, and warmed the atmosphere at the cloud
top as shown by thermal infrared data. They suggest that the NEB
expansion may have been initiated around October 2014, when bulges of
dark colors appeared on the northside of the NEB (16–-18$^\circ$N).
This expansion only extended half way around the planet, and the NEB
had returned to its normal state by June 2016. After the
reorganization of the NTB/NTrZ, a second NEB expansion started in
early 2017, which extended all around the planet within months
(Fletcher et al., 2018).  At the time of the 2013 VLA observations,
which showed no warm NEB northern extension (ledge), the NEB/NTrZ was
fairly quiet. In contrast, in Jan. 2017, ALMA data, which probed
similar pressure levels, did show the ledge during a time that the
NTrZ was highly disturbed and the NEB about to begin an expansion. We
therefore suggest that the presence of the ledge is possibly related
to these large-scale visible cycles in the belts. Moreover, similar to
the 2017 ALMA data with the ledge, 1.3-cm VLA maps from December 2014,
when the NEB was likewise disturbed preceding an expansion, also
showed a broader NEB profile than the 2--4 cm data taken earlier that
year (Fig. 6 in dP19), corroborating our hypothesis.

\subsection{Moist Convection in the SEB}

ALMA observed Jupiter just a few days after an outbreak, or a bright white plume, was reported in the SEB. The spot appeared on 29 December 2016 at a jovigraphic latitude of 16.5$^\circ$S and System-II longitude of 208$^\circ$ (equivalent to System-III longitude of 300.8$^\circ$), coincident with a small white vortex, likely a cyclonic region given the latitudinal gradient in windshear (Fig.~\ref{fig:S1}). Over the next few months new white spots kept appearing within a few degrees of the same System-II longitude (i.e., at a fixed position on Jupiter's disk), while the spots expanded northward producing increasingly extended rifts or disturbances towards the east, i.e., in the prograde direction propelled along by the winds and strong wind shear at those latitudes (Mizumoto, 2017; Rogers, 2018).
The event shows a strong resemblance with the SEB revival in 2010-2011
(Fletcher et al., 2017a), a series of convective events that followed
a period (in 2009--2010) during which the SEB was in a faded state
(Fletcher et al., 2011). Although, as mentioned in the Introduction,
this most recent event was not preceded by an overall fading, the outbreak in both cases was initiated by a series of convective eruptions at a cyclonic spot.

\begin{figure*}
\includegraphics[width=32pc]{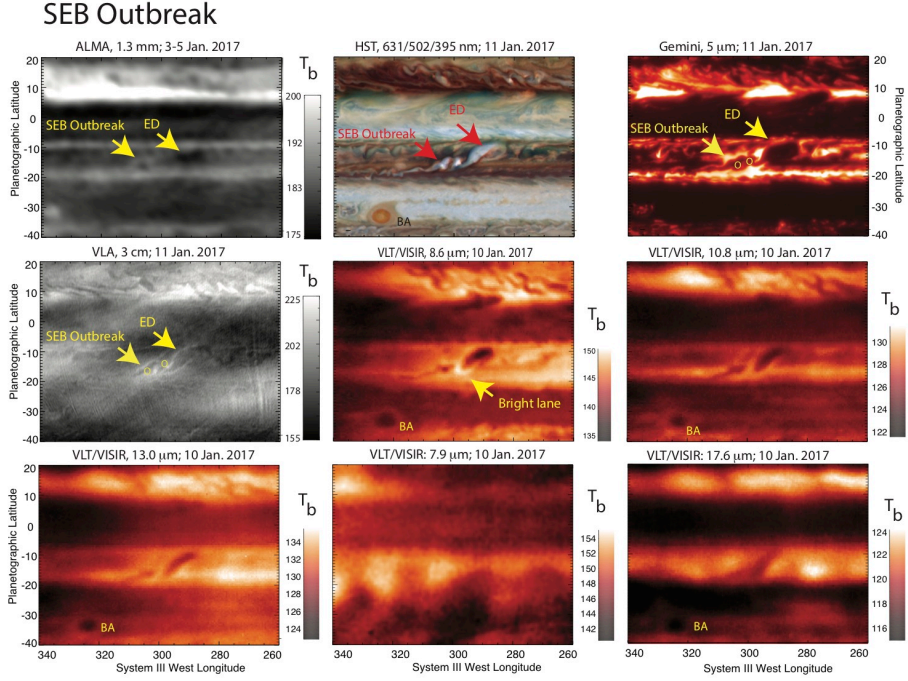}
\caption{The SEB outbreak as observed at different wavelengths. Each panel is about 80$^\circ$ in longitude and 60$^\circ$ in latitude.  The source of the outbreak, as well as the ED (East Disturbance) is indicated in several panels. The various wavelengths are sensitive to different gases and/or aerosols, and hence probe different depths (and hence different temperatures).
}
\label{fig:SEBoutbreaka}\end{figure*}

\begin{figure*}
\includegraphics[width=32pc]{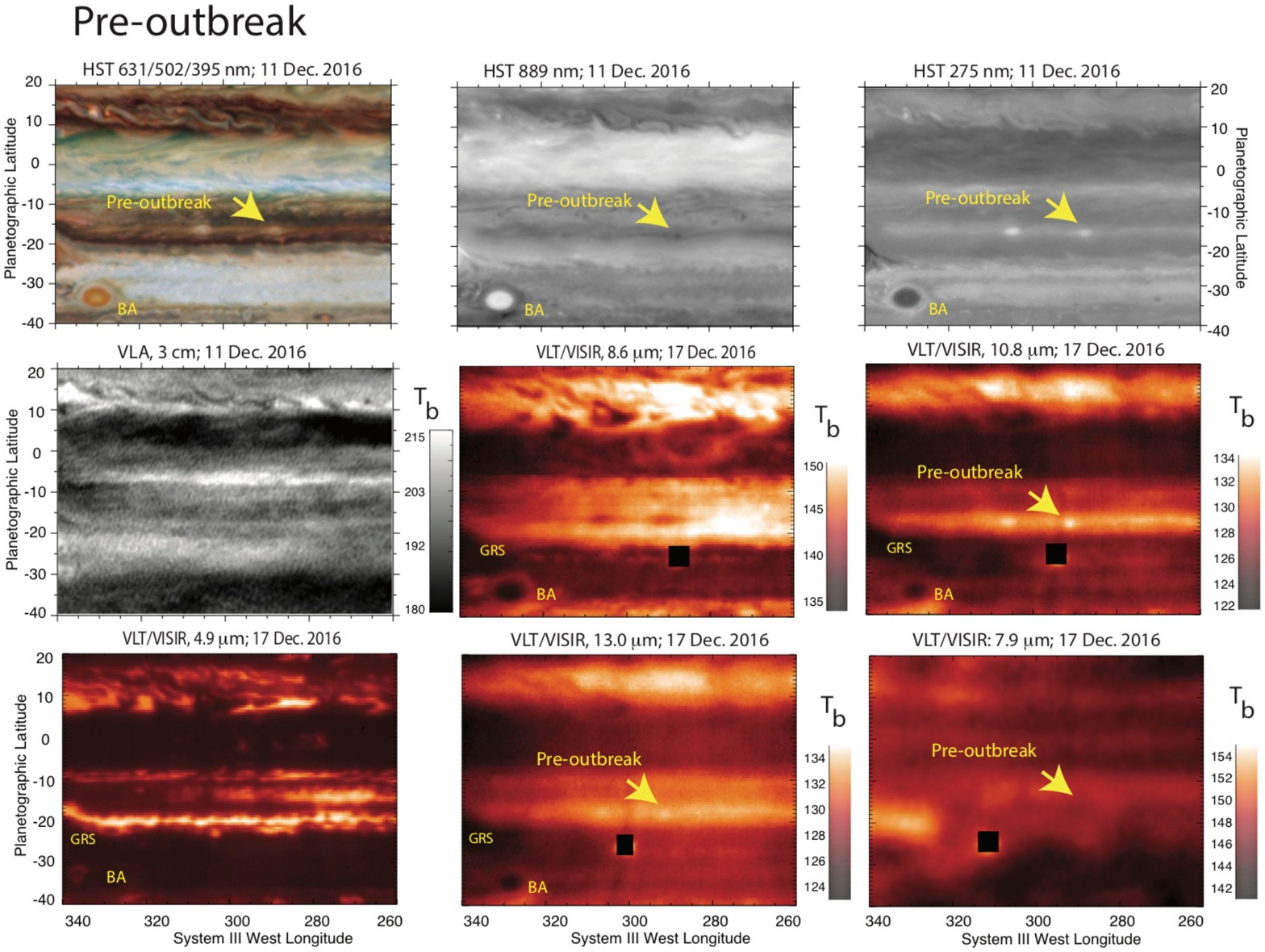}
\caption{The site of the SEB outbreak in HST and VLA maps (11
  Dec. 2016) and at mid-infrared wavelengths (17 Dec. 2016) several weeks before the first plume was seen. The vortex (bright in the visible and mid-IR; dark at 889 nm) where the outbreak took place is indicated by an arrow. The dark square on the mid-IR images blocks out Io, which was visible (saturated) in a background frame. 
}
\label{fig:SEBoutbreakb}\end{figure*}

In contrast to the 2010-2011 SEB revival, the present outbreak was also observed in the mm--cm wavelength range, i.e., including wavelengths that probe {\it below} the cloud layers. Figure~\ref{fig:SEBoutbreaka} 
shows a compilation of images featuring the SEB outbreak in January 2017 at different wavelengths. Figure~\ref{fig:SEBoutbreakb} shows the same region several weeks earlier, in December 2016. The pre-outbreak spot is bright in reflected-light visible and UV HST images, indicative of aerosols; the south side of the spots is warm, as shown by the thermal infrared 10.8 and 13 $\mu$m VLT images. At 889 nm, in the methane absorption band, the spot reveals a small dark center, which implies that aerosols in the upper part of the cyclone must be small ($\lesssim$0.1 $\mu$m) to be reflective in the UV but transparent at 889 nm. At 5 and 8.7 $\mu$m the area at and around the spot is dark, indicative of clouds that prevent deeper-seated emission from leaking through. 
No disturbance is seen at radio wavelengths, where the main source of opacity is NH$_3$ gas, and clouds/hazes are transparent.

The pre-outbreak spot is very similar to the one $\sim$20$^\circ$ to the west, except that it displays the small clearing at 889 nm -- it is not clear how this difference could predict such a vigorous eruption a few weeks later.

The source of the SEB outbreak (Fig.~\ref{fig:SEBoutbreaka}) is dark in the ALMA map, surrounded by a brighter ring, indicative of NH$_3$ gas rising up to higher (colder) altitudes, with dry gas subsiding around the periphery, like the secondary circulation in small vortices (de Pater et al., 2010). A small brighter lane is visible to the northeast, connecting to a large dark area, referred to as the East Disturbance (ED). The brightness temperature at 1.3 mm of both the source of the SEB outbreak and the ED is consistent with a model of NH$_3$ gas rising up from the deep atmosphere (Fig.~\ref{fig:modelfits2}). 
 
At mid-IR wavelengths, data taken $\sim$6 days after the ALMA observations, the SEB outbreak and ED are dark at all troposphere-sensing wavelengths (8.6 -- 20 $\mu$m), indicative of cold temperatures, enhanced aerosol opacity, or (most likely) a combination of the two (Fig.~\ref{fig:SEBoutbreaka}). 
VLA, HST, and 5-$\mu$m observations were taken one day later.  At this time two prominent convective storms are visible on the HST map, with the ED to the northeast. The location of both plumes is indicated on the VLA and 5-$\mu$m maps (yellow circles in the dark areas), both indicative of low brightness temperatures. The ED is also dark on these maps, while bright regions near/around the plume locations and along the ED periphery imply aerosol-free dry subsiding air, so deeper warmer layers are probed. The bright lane 
 at the southwestern edge of the ED in the 8.6-$\mu$m image, sensing a combination of temperature and aerosol opacity at the 500-mbar level, is consistent with this picture. This lane had moved slightly northwards between January 10 and 11, as shown in the Gemini image (Fig.~\ref{fig:SEBoutbreaka}); on 10 January, the 5-$\mu$m region coincided in position with the 8.6-$\mu$m lane (not shown). 
At higher altitudes sensed by the 17.6/18.7/19.5-$\mu$m images, the SEB outbreak and ED are simply cold and embedded in the warmer SEB; due to the lower spatial resolution, details of the structure are washed out.
A tail to the south-west of the SEB outbreak is visible in all images, the direction of which is consistent with the gradient in the wind profile. 
Finally,  at 7.9 $\mu$m, probing the stratosphere, a wave with a 20-25 degree longitudinal spacing might be present towards the west of the source outbreak. Such a stratospheric thermal wave was clearly present during the 2010-2011 SEB revival (Fletcher et al., 2017a).

From all maps together, we infer that NH$_3$ gas is most likely
brought up in the convective plume(s), drying out through
condensation, and descending along the periphery. Model fits to the
plume at the source of the SEB outbreak in the HST data (Section~\ref{sec:rt_res2}) corroborate this picture. 
The cumulative opacity in the SEB plume is about twice that of the
background at high altitudes (100--200 mbar), with larger-sized particles, and a quarter of the cumulative opacity between 0.2--0.7 bar
(Fig.~\ref{fig:hstmodel}), such as would be expected if particles rise
up to much higher altitudes in the plume region.  
This suggests that the plume consists of particles
thrown from lower altitudes high into the atmosphere.

Simultaneously with the Gemini images, we took 5-$\mu$m spectra near
the SEB outbreak using NIRSPEC at Keck (Fig.~\ref{fig:niri}). These
spectra were taken very close to a bright (hot spot) area, and we have
good reasons to believe the data are contaminated by flux from these
hot regions. Nevertheless, we can conclude from the in-depth analysis
in Section~\ref{sec:rtkeck} that the spectra 
are consistent with the ED and plume region having thick clouds at the 600-mbar level. Although a best fit to the spectrum suggests no cloud at the 5-bar level, we do not trust this. The gas composition is similar to that of adjacent Hot Spots, with H$_2$O at 47 ppm at $P > 4.5$ bar (as measured with the Galileo Probe; Wong et al., 2004), and zero at higher altitudes. NH$_3$ line profiles were best matched using 300 ppm at $P > 1.2$ bar. With the likely contamination by hot spots, this NH$_3$ abundance should be taken as a lower limit. Although our ALMA data agree well with this abundance (Fig.~\ref{fig:modelfits}), due to ALMA's low spatial resolution this also was taken as a lower limit to the NH$_3$ abundance.


Fletcher et al. (2017a) compared the convective eruptions triggering the 2010-2011 SEB revival with mesoscale convective storms (MCS) seen on Earth, which show intense precipitation and cold cloud tops (Houze, 1993). As discussed above, the SEB eruptions, like those in the NTB, are probably moist-convective plumes, rising up from the water condensation level. While injection of energy warms the atmosphere relative to its surroundings, resulting in a cyclonic motion, near the top of the plume, where divergence and cooling takes place, an anticyclonic motion is expected (Emanuel, 1994). Since such a motion is in the opposite sense to that expected from the windshear across the SEB, the anticyclones will not persist for long but break up into eddies. Such a sequence of events, starting with a moist-convective plume and ending with its demise, while new plumes arise at the same location (same System-II longitude) was imaged at high spatial resolution by Voyager 1 in Feb. 1979, and modeled by Hueso et al. (2002). The Voyager images closely resemble the present, as well as previous, SEB outbreaks, including the eruption, westward tail, and ED. None of the previous observations, however, yields information below the visible ammonia clouddeck. Our ALMA observations are the first to show that high concentrations of NH$_3$ gas are brought up in the plume, i.e., the source of the outbreak, as well as in the disturbance to the east. The mid-infrared images show that the top of the plumes are indeed cold, as expected from the models. Hence our data are fully consistent with models of moist convection.

\section{Summary}

This paper focuses on 1.3- and 3-mm maps constructed from data obtained with ALMA on 3--5 January 2017, just days after the onset of an outbreak in the SEB, and a few months after a reorganization of the NTB. These data are the first to characterize the atmosphere {\it below} the cloud layers during/following such outbreaks. 
Aided also by observations ranging from uv to mid-infrared
wavelengths, we have shown that the eruptions are consistent with
models where energetic plumes are triggered via moist convection at
the base of the water cloud. The plumes bring up ammonia gas from the
deep atmosphere to high altitudes, where NH$_3$ gas is condensing out
and the subsequent dry air is descending in neighboring regions. The
cloud tops are cold, as shown by mid-infrared data, indicative of an
anticyclonic motion, which causes the storm to break up, as expected
from similarities to mesoscale convective storms on Earth. The plume
particles reach altitudes as high as the tropopause.

Our research shows the importance of simultaneous multi-wavelength
observations of transient events, that sense the atmosphere from below the cloud layers to well above the tropopause.

\section*{Acknowledgements}

This research was supported by NASA's Planetary Astronomy (PAST) award
NNX14AJ43G and Solar System Observations (SSO) award 80NSSC18K1001 to
the University of California, Berkeley. CM was supported in part by
the NRAO Student Observing Support (SOS) Program. MW and GB were
supported in part by Solar System Observations (SSO) award SSO NNX15AJ41G.
LF was supported by a Royal Society Research Fellowship and European
Research Council Consolidated Grant at the University of Leicester. JS
and RC were supported by NASA Postdoctoral Fellowships. GO and JS were also supported by a contract between the Jet Propulsion Laboratory/California Institute of Technology and NASA. We thank Andrew S. Wetzel (Clemson University) for his help in reducing the Keck/NIRSPEC data.

This paper makes use of ALMA data 2016.1.00701.S, and VLA data VLA/16B-048.  ALMA is a partnership of ESO (representing its member states), NSF (USA) and NINS (Japan), together with NRC (Canada), MOST and ASIAA (Taiwan), and KASI (Republic of Korea), in cooperation with the Republic of Chile. The Joint ALMA Observatory is operated by ESO, AUI/NRAO and NAOJ. The data can be downloaded from the ALMA Archive. The National Radio Astronomy Observatory is a facility of the National Science Foundation operated under cooperative agreement by Associated Universities, Inc.

This research was partially based on thermal-infrared observations acquired at the ESO Very Large Telescope (VLT) Paranal UT3/Melipal Observatory (098.C-0681(C) and 098.C-0681(D)); all data are available via the ESO science archive.

The research was also in part based on Gemini data (GN-2016B-FT-18). The Gemini observatory is operated by the Association of Universities for Research in Astronomy, Inc., under a cooperative agreement with the NSF on behalf of the Gemini partnership: the National Science Foundation (United States), the National Research Council (Canada), CONICYT (Chile), the Australian Research Council (Australia), Minist\'erio da Ci\^encia, Tecnologia e Inovac\"ao (Brazil) and Ministerio de Ciencia, Tecnolog\'ia e Innovaci\'on Productiva (Argentina). 

We further used observations (GO 14839 and GO-14661) made with the NASA/ESA Hubble Space Telescope (HST) at the Space Telescope Science Institute, which is operated by the Association of Universities for Research in Astronomy, Inc., under NASA contract NAS 5-26555, with support provided by NASA through a grant from the Space Telescope Science Institute. 

COMICS images were obtained at the Subaru telescope, which is operated by the National Astronomical Observatory of Japan (NAOJ). 
Part of these data were awarded through the Keck-Subaru time exchange program. NIRSPEC data were acquired with the Keck 2 telescope (2016B$_-$N045NS). The W. M. Keck Observatory is operated as a scientific partnership among the California Institute of Technology, the University of California and NASA (the National Aeronautics and Space Administration) and supported by generous financial support of the W. M. Keck Foundation.

The authors also wish to recognize and acknowledge the very significant cultural role and reverence that the summit of Maunakea has always had within the indigenous Hawaiian community. We are most fortunate to have the opportunity to conduct observations from this mountain.  

\facilities{ALMA, VLA, HST(WFC2/UVIS), Keck, Gemini, VLT, Subaru}

\pagebreak

\section*{References}

\'Ad\'amkovics, et al. 2016. Meridional variation in tropospheric methane on Titan observed with AO spectroscopy at Keck and VLT, Icarus, 270, 376-388.

Asay-Davis X.S., Marcus, P.S., Wong, M. H.,  de Pater, I.,  2011.
Changes in Jupiter's Zonal Velocity between 1979 and 2008.
Icarus,  211, 1215-1232.

Atkinson, D.H., Pollack, J.B., Seif, A., 1998. The Galileo probe doppler wind experiment: measurement of the deep zonal winds on Jupiter. J. Geophys. Res. 103 (E10), 22911-22928.

Bjoraker, G.L., Wong, M.H., de Pater, I., Hewagama, T., \'Ad\'amkovics, M., Orton, G.S., 2018. The Gas Composition and Deep Cloud Structure of Jupiter's Great Red Spot. Astron. J., 156, \#101, 15 pp.
   

Brown, S., et al., 2018. Prevalent lightning sferics at 600 megahertz near Jupiter's poles. Nature, 558, 87-90 (DOI: 10.1038/s41586-018-0156-5)

Conrath, B.J., Gierasch, P.J., Leroy, S.S., 1990. Temperature and circulation in the stratosphere
of the outer planets. Icarus, 83, 255-281.




 de Kleer, K., Luszcz-Cook, S., de Pater, I., Adamkovics, M.,
    Hammel, H., 2015. Clouds and aerosols on Uranus: Radiative
    transfer modeling of spatially-resolved near-infrared Keck
    spectra. Icarus, 256, 120-137.

de Pater, I., 1986. Jupiter's zone-belt structure at radio wavelengths: 
  II. Comparison of observations with model atmosphere calculations, 
  Icarus,  68, 344-365.

 de Pater, I., and D.L. Mitchell, 1993, Microwave Observations of the 
Planets: the Importance of Laboratory Measurements,  J. Geophys. Res. 
Planets, 98, 5471-5490. 

 de Pater, I., D. Dunn, K. Zahnle and P.N. Romani,
2001. Comparison of Galileo Probe Data with Ground-based Radio
Measurements. Icarus, 149, 66-78

de Pater, I., DeBoer, D.R., Marley, M., Freedman, R.,  Young, R.,
2005. Retrieval of water in Jupiter's deep atmosphere using
microwave spectra of its brightness temperature. Icarus, 173, 425-438.

de Pater, I., Wong, M. H., Marcus, P. S., Luszcz-Cook, S., \'Ad\'amkovics, M., 
Conrad, A., Asay-Davis, X., Go, C., 2010.
   Persistent Rings in and around Jupiter's Anticyclones - Observations and Theory.
Icarus, 210, 742-762.

de Pater, I., Wong, M. H., de Kleer, K., Hammel, H. B., \'Ad\'amkovics, M.,
Conrad, A., 2011. 
Keck Adaptive Optics Images of Jupiter's North Polar Cap and Northern Red Oval. Icarus, 213, 559-563.

de Pater, I.,  Fletcher, L.N.,  Luszcz-Cook, S.H., DeBoer, D.,
    Butler, B., Hammel, H.B., Sitko, M.L., Orton, G.,  Marcus, P.S., 2014. Neptune's Global
Circulation deduced from Multi-Wavelength Observations. Icarus,
    237, 211-238.

de Pater, I., Sault, R. J., Butler, B., DeBoer, D., Wong, M. H., 2016. Peering through Jupiter's Clouds with Radio Spectral Imaging, Science, 352, Issue 6290, pp. 1198-1201. (referred to as dP16).
 
de Pater, I., Sault, R. J.,  Wong, M. H., Fletcher, L. N., DeBoer, D.,  Butler, B.,  2019. Jupiter's ammonia distribution derived from VLA maps at 3--37 GHz. Icarus, 322, 168-191. (referred to as dP19)

Dressel, L. Wide Field Camera 3 Instrument Handbook, version 11.0 (STScI, Baltimore MD, 2019).

Emanuel K 1994 Atmospheric Convection (New York: Oxford University Press)

Fletcher, L. N., 2017. Cycles of activity in the Jovian atmosphere. Geophys. Res. Lett. 44, 4725-4729.

Fletcher, L. N., G. S. Orton, P. Yanamandra-Fisher, B. M. Fisher, P. D. Parrish, and P. G. J. Irwin, 2009. Retrievals of atmospheric variables on the gas giants from ground‐based mid‐infrared imaging, Icarus, 200, 154–175, doi:10.1016/j.icarus.2008.11.019.
 
Fletcher, L.N., G.S. Orton, J.H. Rogers, A. A. Simon-Miller, I. de Pater, M.H. Wong, O. Mousis, P.G.J. Irwin, M. Jacquesson, P.A. Yanamandra-Fisher, 2011. Jovian Temperature and Cloud Variability during the 2009-2010 Fade of the South
Equatorial Belt. Icarus, 213, 564-580.

Fletcher, L. N., Greathouse, T. K., Orton, G. S., Sinclair, J. A., Giles, R. S., Irwin, P. G. J., Encrenaz, T., 2016. Mid-infrared mapping of Jupiter's temperatures, aerosol opacity and chemical distributions with IRTF/TEXES. Icarus, 278, p. 128-161. 

Fletcher, L. N., Orton, G. S.,  Rogers, J. H., Giles, R. S., Payne, A. V., Irwin, P. G. J., Vedovato, M., 2017a. Moist Convection and the 2010-2011 Revival of Jupiter’s South Equatorial Belt. Icarus, 286, 94--117.

Fletcher, L. N., Orton, G. S., Sinclair, J. A., et al., 2017b. Jupiter's North Equatorial Belt expansion and thermal wave activity ahead of Juno's arrival, Geophys. Res. Lett., 44, 7140-7148.


Fletcher, L. N.,  et al., 2018.  Jupiter's Mesoscale Waves Observed at 5 $\mu$m by ground-based observations and Juno JIRAM.  Astronomical Journal, 156, 67-80. 


Gibson, J., Welch, Wm. J.,  de Pater, I., 2005. Accurate Jovian
Flux Measurements at $\lambda$1cm Show Ammonia to be Sub-saturated in
the Upper Atmosphere. Icarus, 173, 439-446.

Hodapp, K. W., et al., 2003. The Gemini Near-Infrared Imager (NIRI). Publ. Astron. Soc. Pac., 115, 814.

Houze, R., 1993. Cloud Dynamics. Vol. 53 of International Geophysics Series.
Academic Press.

Hueso, R., S\'anchez-Lavega, A., Guillot, T., Oct. 2002. A model for large-scale Moist Convection for the Giant Planets: The Jupiter Case. Icarus 151, 257–274.

 Irwin, P. G., Bowles, N., Braude, A. S., Garland, R., \& Calcutt, S., 2018. Analysis of gaseous
ammonia (NH$_3$) absorption in the visible spectrum of Jupiter, Icarus, 302, 426-436.


Karim, R. L., deBoer, D., de Pater, I., Keating, G. K., 2018. A Wideband Self-consistent Disk-averaged Spectrum of Jupiter Near 30 GHz and Its Implications for NH$_3$ Saturation in the Upper Troposphere. Astron. J., 155, article id. 129, 8 pp.

Kataza et al., 2000, COMICS: the cooled mid-infrared camera and spectrometer for the Subaru telescope, Proceedings of Society of Photo-Optical Instrumentation Engineers (SPIE), Vol. 4008, 1144–1152

Kunde, et al., 1996, Cassini infrared Fourier spectroscopic investigation, Proceedings of Society of Photo-Optical Instrumentation Engineers (SPIE), volume 280, 162–177. 

Lagage, P. O., et al. (2004), Successful commissioning of VISIR: The mid-infrared VLT instrument, Messenger, 117, 12-16.
 
Li, C., et al., 2017. The distribution of ammonia on Jupiter from a preliminary
inversion of Juno microwave radiometer data. Geophys. Res. Lett., 44, 5317–5325, doi:10.1002/2017L073159.

Lii, P.S., M.H. Wong, and I. de Pater, 2010. Temporal Variation of the 
  Tropospheric Cloud and Haze in the Jovian Equatorial Zone. Icarus, 209, 591-601.

 Lindal, G.F., 1992, The atmosphere of Neptune: An analysis of radio 
occultation data acquired with Voyager 2, Astron. J., 103, 967-982.

 Luszcz-Cook, S.H., K. de Kleer, I. de Pater, M. Adamkovics, H.B. Hammel, 2016. Retrieving Neptune's aerosol properties from Keck OSIRIS observations. I. Dark regions. Icarus, 276, 52-87.

Marcus, P.S., Tollefson, J., Wong, M.H., de Pater, I., 2019. An
Equatorial Thermal Wind Equation: Applications to Jupiter.
Icarus, 324, 198-223.

McLean, I.S., Becklin, E.E., Bendiksen, O., et al. 1998. Design and development of NIRSPEC: a near-
infrared echelle spectrograph for the Keck II telescope. Proc. SPIE 3354, 566-578.

Mizumoto, S., 2017. ‘2016-2017 Mid-SEB Outbreak Final Report’, posted on ALPO-Japan web site: alpo-j.asahikawa-med.ac.jp/kk17/j170923s.htm 

Moeckel, C., Janssen, M.J., de Pater, I., 2019. A fresh look at the Jovian radio emission as seen by
Cassini-RADAR and implications for the time
variability. Icarus, 321, 994-1012.

 Molter, E., et al., 2019. Discovery of a bright equatorial storm on Neptune. Icarus, 321, 324-345.

Rogers, J. H., 1995. The Giant Planet Jupiter, 418 pp., Cambridge Univ. Press, Cambridge, U. K.

Rogers, J. H., 2018. Jupiter in 2016-17, Report no.17: ‘Summary of the mid-SEB outbreak’, on the BAA web site:  https://www.britastro.org/node/16772


S\'anchez-Lavega, et al., 2017.  A planetary-scale disturbance in the most intense Jovian atmospheric jet from JunoCam and ground-based observations.  Geophysical Res. Lett.  44, 4679-4686.


Sault, R.J., Engel, C., de Pater, I., 2004. Longitude-resolved Imaging
of Jupiter at $\lambda=2$ cm. Icarus, 168, 336-343.

Sault, R. J., Teuben, P. J., \& Wright, M. C. H. 1995, A Retrospective View of MIRIAD, in ASP Conf. Ser. 77,
Astronomical Data Analysis Software and Systems IV, ed. R. A. Shaw,
H. E. Payne, \& J. J. E. Hayes (San Francisco: ASP), 433-436.

Showman, A.P., de Pater, I., 2005. Dynamical implications of Jupiter's
tropospheric ammonia abundance. Icarus, 174, 192-204.

Sugiyama, K., Nakajima, K., Odaka, M., Kuramoto, K., Hayashi, Y.-Y., 2014. Numerical simulations of Jupiter’s moist convection layer: Structure and dynamics in statistically steady states. Icarus, 229, 71-91.

Tollefson, J., Wong, M.H.,  de Pater, I.,  Simon, A., Orton, G.S.,
Rogers, J.H.,  Atreya, S.K,  Cosentino, R.G., Januszewski, W.,
Morales-Juberias, R., Marcus, P.S., 2017. Changes in Jupiter's Zonal Wind Profile preceding and during the Juno mission. Icarus, 296, 163-178.

van Dokkum, P. G., 2001. Cosmic-ray rejection by Laplacian edge
detection. Publ. Astron. Soc. Pacific, 113, 1420-1427. 

Vasavada, A. R., Showman, A. P., 2005. Jovian atmospheric dynamics: an update after Galileo
and Cassini. Rep. Prog. Phys. 68, 1935–1996.

Wong, M.H., Mahaffy, P.R., Atreya, S.K., Niemann, H.B., Owen,  T.C., 2004. Updated Galileo probe mass spectrometer measurements of carbon, oxygen, nitrogen, and sulfur on Jupiter. Icarus 171, 153-170. 

Wong, M.H., 2011. Fringing in the WFC3/UVIS detector. In Proc. 2010 Space Telescope Science Institute Calibration Workshop (eds Deustua, S. \& Oliveira, C.) 189-200 (STScI, Baltimore MD, 2011).

Wong, M. H., Atreya, S. K., Kuhn, W. R., Romani, P. N., \& Mihalka, K. M.
2015. Fresh clouds: A parameterized updraft method for calculating
cloud densities in one-dimensional models. Icarus, 245, 273




Wong, M.H., et al., 2019. High-resolution UV/optical/IR imaging of
Jupiter in 2016-2018. In Prep.


\end{document}